\documentclass[manuscript,screen]{acmart}
\AtBeginDocument{%
  }

\setcopyright{acmlicensed}
\copyrightyear{2024}
\acmYear{2024}
\acmDOI{XXXXXXX.XXXXXXX}
\acmISBN{978-1-4503-XXXX-X/18/06}

\usepackage{xcolor}
\usepackage{xspace}
\usepackage{tabularx}% http://ctan.org/pkg/tabularx
\usepackage{colortbl}% http://ctan.org/pkg/colortbl
\usepackage{url}
\usepackage{hyperref}
\usepackage{graphicx}
\usepackage{array}			% align vertically in tables
\usepackage{multirow}
\usepackage{tabularx}
\usepackage{booktabs} % Horizontal rules in tables

\usepackage{tikz}

\usepackage{ifthen}
\newboolean{showcomments}
\setboolean{showcomments}{true} % toggle to show or hide comments
\ifthenelse{\boolean{showcomments}}
{\newcommand{\nb}[2]{
    \fcolorbox{gray}{yellow}{\bfseries\sffamily\scriptsize#1}
    {$\blacktriangleright$#2$\blacktriangleleft$}
  }
  
}
{\newcommand{\nb}[2]{}
  
}

\newcommand\obs[1]{\nb{OBS}{\textcolor{brown}{\textsl{#1}}}}

\newcommand{\oapp}{CoRAL\xspace}
\usepackage{subcaption}

\newcommand{\eg}{e.g.,~}							% exempli gratia (for the sake of example)
\newcommand{\ie}{i.e.,~}							% id est (that is)
\newcommand{\etal}{~et al.}					% et alia (and others)
\newcommand{\Fig}[1]{Figure~\ref{#1}}  			% choose Fig. or Figure, depending on the style
\newcommand{\Table}[1]{Table~\ref{#1}}	    % Table reference
\newcommand{\Sect}[1]{Section~\ref{#1}}	  % section name always with a capital S
\usepackage{tcolorbox}
\usepackage{listings}
\lstdefinestyle{mystyle}{
    backgroundcolor=\color{gray!10},
    commentstyle=\color{gray!80},
    keywordstyle=\bfseries\color{red},
    numberstyle=\tiny\color{gray},
    stringstyle=\color{blue},
    basicstyle=\ttfamily\footnotesize,
    breakatwhitespace=false,
    breaklines=true,
    captionpos=b,
    keepspaces=true,
    numbers=left,
    numbersep=5pt,
    showspaces=false,
    showstringspaces=false,
    showtabs=false,
    tabsize=2
}

\usepackage{listings}
\usepackage{tcolorbox}
\usepackage{xcolor}

\lstdefinestyle{mystyle}{
    basicstyle=\ttfamily\small,
    keywordstyle=\color{blue},
    commentstyle=\color{gray},
    stringstyle=\color{red},
    showstringspaces=false,
    breaklines=true,
    frame=tb,
    framerule=0.5pt,
    framesep=3pt,
    xleftmargin=12pt
}

\tcbset{
    codebox/.style={
        colframe=black!50,
        colback=white,
        sharp corners,
        boxrule=0.5pt,
        top=8pt,
        bottom=8pt
    }
}

\begin{document}

%%
%% The "title" command has an optional parameter,
%% allowing the author to define a "short title" to be used in page headers.
\title{Leveraging Reward Models for Guiding Code Review Comment Generation}

\thanks{The replication package is available online \cite{github_replication}.}
%%
%% The "author" command and its associated commands are used to define
%% the authors and their affiliations.
%% Of note is the shared affiliation of the first two authors, and the
%% "authornote" and "authornotemark" commands
%% used to denote shared contribution to the research.
\author{Oussama Ben Sghaier}
\email{trovato@corporation.com}
\orcid{0000-0003-2737-0952}
\affiliation{%
  \institution{Université de Montréal}
  \country{Canada}
}

\author{Rosalia Tufano}
\email{trovato@corporation.com}
\orcid{0009-0009-7017-3066}
\affiliation{%
  \institution{Università della Svizzera italiana}
  \country{Switzerland}
}

\author{Gabriele Bavota}
\email{trovato@corporation.com}
\orcid{0000-0002-2216-3148}
\affiliation{%
  \institution{Università della Svizzera italiana}
  \country{Switzerland}
}

\author{Houari Sahraoui}
\email{sahraouh@iro.umontreal.ca}
\orcid{0000-0001-6304-9926}
\affiliation{%
  \institution{Université de Montréal}
  \country{Canada}
}

%%
%% By default, the full list of authors will be used in the page
%% headers. Often, this list is too long, and will overlap
%% other information printed in the page headers. This command allows
%% the author to define a more concise list
%% of authors' names for this purpose.
\renewcommand{\shortauthors}{Ben Sghaier et al.}

\begin{abstract}
Code review is a crucial component of modern software development, involving the evaluation of code quality, providing feedback on potential issues, and refining the code to address identified problems. Despite these benefits, code review can be rather time consuming, and influenced by subjectivity and human factors. For these reasons, techniques to (partially) automate the code review process have been proposed in the literature. Among those, the ones exploiting deep learning (DL) are able to tackle the generative aspect of code review, by commenting on a given code as a human reviewer would do (i.e., comment generation task) or by automatically implementing code changes required to address a reviewer's comment (i.e., code refinement task).
In this paper, we introduce \oapp, a deep learning framework automating review comment generation by exploiting reinforcement learning with a reward mechanism considering both the semantics of the generated comments as well as their usefulness as input for other models automating the code refinement task. The core idea is that if the DL model generates comments that are semantically similar to the expected ones or can be successfully implemented by a second model specialized in code refinement, these comments are likely to be meaningful and useful, thus deserving a high reward in the reinforcement learning framework.
We present both quantitative and qualitative comparisons between the comments generated by \oapp and those produced by the latest baseline techniques, highlighting the effectiveness and superiority of our approach.

\end{abstract}

%%
%% The code below is generated by the tool at http://dl.acm.org/ccs.cfm.
%% Please copy and paste the code instead of the example below.
%%
\begin{CCSXML}
<ccs2012>
 <concept>
  <concept_id>00000000.0000000.0000000</concept_id>
  <concept_desc>Do Not Use This Code, Generate the Correct Terms for Your Paper</concept_desc>
  <concept_significance>500</concept_significance>
 </concept>
 <concept>
  <concept_id>00000000.00000000.00000000</concept_id>
  <concept_desc>Do Not Use This Code, Generate the Correct Terms for Your Paper</concept_desc>
  <concept_significance>300</concept_significance>
 </concept>
 <concept>
  <concept_id>00000000.00000000.00000000</concept_id>
  <concept_desc>Do Not Use This Code, Generate the Correct Terms for Your Paper</concept_desc>
  <concept_significance>100</concept_significance>
 </concept>
 <concept>
  <concept_id>00000000.00000000.00000000</concept_id>
  <concept_desc>Do Not Use This Code, Generate the Correct Terms for Your Paper</concept_desc>
  <concept_significance>100</concept_significance>
 </concept>
</ccs2012>
\end{CCSXML}

\ccsdesc[500]{Do Not Use This Code~Generate the Correct Terms for Your Paper}
\ccsdesc[300]{Do Not Use This Code~Generate the Correct Terms for Your Paper}
\ccsdesc{Do Not Use This Code~Generate the Correct Terms for Your Paper}
\ccsdesc[100]{Do Not Use This Code~Generate the Correct Terms for Your Paper}

%%
%% Keywords. The author(s) should pick words that accurately describe
%% the work being presented. Separate the keywords with commas.
\keywords{Code review, reinforcement learning, code analysis, software maintenance.}

\received{20 February 2007}
\received[revised]{12 March 2009}
\received[accepted]{5 June 2009}

%%
%% This command processes the author and affiliation and title
%% information and builds the first part of the formatted document.
\maketitle

\section{Introduction}
\label{sec:introduction}

Code review is a crucial part of the software development lifecycle, and aims at identifying issues, suboptimal implementation choices, and bugs \cite{mcintosh2014impact, mcintosh2016empirical} ensuring the overall quality of the source code \cite{ackerman1989software, morales2015code}. This process primarily involves one or more developers manually examining the code written by their peers. Key tasks in code review include estimating the quality of submitted code, identifying potential issues through review comments, and refining the code to address these issues.

Numerous studies in the literature highlight several benefits of code review, including higher quality code, reduced technical debt, bug prevention, and enhanced knowledge transfer among developers. However, the code review process is often perceived as time-consuming, expensive, and complex, especially in large-scale projects \cite{eick2001does, avgeriou2016managing} where thousands of code reviews may occur \cite{bosu2013impact}. Additionally, this process is notably subjective, influenced by various human and social factors such as the experience levels of developers and their interpersonal relationships. This can introduce biases, leading to inefficiencies and inconsistencies that ultimately impact the overall quality and reliability of the codebase~\cite{ackerman1989software, morales2015code}.

To support reviewers in this challenging task, one possibility is to employ static analysis to automatically identify potential issues \cite{pmd, findBugs}. These tools utilize manually-defined rules to establish standards and highlight code fragments that violate them. However, the efficacy of these tools is limited, as their rules require constant updates to address a broad spectrum of emerging issues. Moreover, software problems evolve over time and can be influenced by factors such as architecture, team culture, and employed technologies. Therefore, the inflexible nature of static analysis tools diminishes their utility and effectiveness in dynamic and evolving software projects~\cite{bielik2017learning, sadowski2015tricorder}.

Alternative methods use similarity techniques to suggest relevant review comments from a predefined dataset based on similarities to code changes \cite{siow2020core, gupta2018intelligent}. Although these recommendations can be beneficial, review comments are often context-specific rather than generic and, thus, reusable.

Recent advancements in deep learning (DL) and natural language processing have sparked interest in leveraging pre-trained language models to automate various software engineering tasks. In particular, recent works \cite{tufano2022using, tufan2021towards, li2022automating, sghaier2023multi} have explored the application of generative AI, specifically language models, to automate code review tasks, such as the generation of review comments as a human reviewer would do (i.e., posting comments to report quality issues in the reviewed code), or the automatic implementation of review comments (i.e., code refinement task). Although these methods have shown promising results, code review tasks have traditionally been addressed in isolation, without considering their significant interdependencies. To address this limitation, Sghaier \etal \cite{sghaier2024improving} proposed an approach based on cross-task knowledge distillation to simultaneously address comment generation and code refinement. Despite the improvement in quality brought by this work over the state of the art, the quality of the generated comments remains suboptimal as it relies only on a simple combination of loss functions. In our work, we exploit reinforcement learning (RL) to explore a variety of additional feedback signals (\ie rewards) that are more meaningful and informative (\eg semantic similarity with a human-written comment, correctness of the subsequent task). Such an approach is inspired by RL from human feedback that has shown promising results in aligning models with human preferences \cite{bai2022training}.

Our proposed framework, called \oapp, leverages RL to generate code review comments. \oapp employs two reward strategies, namely semantic similarity and subsequent task rewards, to guide the generation process. 
In the comment generation task, the LLM takes the patch (\ie code difference) as input and generates a comment. For the subsequent task reward strategy, the generated comment, along with the patch, is fed into the LLM to produce the necessary code edits. The reward value is determined by measuring the correctness of code edits using metrics such as loss or CrystalBLEU \cite{eghbali2022crystalbleu} to compare the generated code edits with real ones. This strategy aims to generate useful comments that facilitate effective code refinement.
Additionally, we implement a semantic similarity reward strategy, where the reward value is the semantic similarity between the generated and real (\ie human-written) comments. This approach ensures that the generated comments, while potentially phrased differently, convey the same meaning as the real comments, thus maintaining their relevance and usefulness.

To evaluate the effectiveness of \oapp, we conduct both quantitative and qualitative evaluations. In the quantitative evaluation, we compare the different reward strategies within \oapp and benchmark \oapp against baseline and state-of-the-art models, using BLEU scores and accumulated rewards as our primary metrics. For the qualitative evaluation, we employ GPT-4 as a judge to assess the usefulness of comments generated by our proposed framework compared to those produced by the baseline. We also perform statistical tests to assess the significance of our results.

The remainder of this paper is structured as follows. \Sect{sec:framework} introduces our proposed framework. \Sect{sec:design} outlines the research questions and setup of the experiments. \Sect{sec:results} presents the evaluation results. \Sect{sec:threats} discusses the threats to the validity of our findings. \Sect{sec:literature} reviews related work. \Sect{sec:conclusion} concludes.

\section{Background}
\label{sec:background}

\subsection{Code review}

In software development, particularly in continuous integration environments, version control systems such as GitHub play a critical role in enabling collaboration and managing codebase evolution \cite{shahin2017continuous, fowler2006continuous}. Developers work on local copies of the codebase, making changes to implement new features or fix issues. Once changes are complete, they are pushed to a shared repository. Developers may open a pull request to propose merging these changes into the main branch. Reviewers are then assigned to inspect the proposed changes, identify potential issues, and provide feedback to the author. This iterative process (i.e., code review) continues until the changes meet the required standards and are approved for merging into the main codebase.

%Code review is a cornerstone of the software development lifecycle, ensuring the quality and maintainability of the codebase \cite{ackerman1989software, morales2015code}. It involves inspecting code changes by peers \cite{fagan2002design, bavota2015four} to identify bugs, suboptimal code fragments, style violations, and other issues \cite{mcintosh2014impact, mcintosh2016empirical}. 

While code review encompasses multiple tasks—such as \emph{quality estimation}, \emph{comment generation}\footnote{In a manual process, this would be ``comment writing'', but it is referred to as ``comment generation'' in our paper due to the usage of LLMs to generate comments.}, and \emph{code refinement}—this paper focuses primarily on comment generation.

\emph{Comment generation} involves reviewers providing feedback on issues such as bugs, security vulnerabilities, or style violations, as well as suggesting improvements like refactoring or documentation updates. Automating this process can produce objective, consistent feedback. \emph{Code refinement} refers to developers addressing these comments by making necessary changes, which are then resubmitted for review. \emph{Quality estimation} consists of evaluating whether a PR meets merging standards.

\subsection{Reinforcement learning}

RL is a machine learning paradigm where an agent learns to make decisions by interacting with an environment to maximize cumulative rewards \cite{barto2021reinforcement}. The agent takes actions, observes outcomes, and receives feedback in the form of rewards, which guide its learning process. RL has been successfully applied in various domains, including robotics, game playing, and natural language processing.

A recent advancement in RL is Reinforcement Learning from Human Feedback (RLHF), where human feedback is used to shape the reward function, enabling the agent to learn behaviors aligned with human preferences \cite{ziegler2019fine}. RLHF has been particularly effective in fine-tuning large language models (LLMs) for tasks like dialogue generation and summarization, where human preferences are difficult to encode explicitly \cite{ouyang2022training}.

Building on RLHF, Reinforcement Learning from AI Feedback (RLAIF) replaces human feedback with feedback from AI systems or tools, reducing reliance on costly human annotation \cite{lee2023rlaif}. RLAIF leverages pre-trained models or rule-based systems to provide rewards, enabling scalable and efficient training of RL agents. This approach has shown promise in tasks where human feedback is scarce or expensive to obtain, such as code review automation and software testing.
\section{Motivating Examples}
\label{sec:motivation}

We present two key examples to motivate our work and highlight common challenges in automated code review comment generation.

\subsection{Example 1: Non-Actionable Feedback} 
This example demonstrates a critical limitation in automated code review systems: the generation of non-actionable feedback. While such comments can spot issues in code, they fail to provide concrete guidance for improvement, reducing their practical usefulness. Indeed, refining the code based on this comment may be challenging.

\begin{tcolorbox}[
    title=Example 1,
    colframe=black!30,
    colback=white,
    coltitle=black,
    fonttitle=\bfseries,
    sharp corners,
    boxrule=0.5pt,
    top=8pt,
    bottom=8pt
]
\textbf{Initial Code:}
\begin{lstlisting}[language=Python, basicstyle=\ttfamily\small]
+ def fetch_user_data(user_id):
+     user_data = database.get_user(user_id)
+     if user_data is not None:
+         return user_data
+     else:
+         return {}
\end{lstlisting}

\textbf{Potential generated Comment:} 
``The code is unnecessarily complex and not well-written.''

\textbf{Expected refined code:}
\begin{lstlisting}[language=Python, basicstyle=\ttfamily\small]
def fetch_user_data(user_id):
-     user_data = database.get_user(user_id)
-     if user_data is not None:
-         return user_data
-     else:
-         return {}
+     return database.get_user(user_id) or {}
\end{lstlisting}
\end{tcolorbox}

\noindent\textbf{Challenge 1.} The generated comment correctly identifies that the code is unnecessarily complex (lines 2–6 in the initial code) but fails to provide actionable guidance for improvement. Such comments, though technically accurate, are unhelpful in practice because they do not offer concrete suggestions for resolving the issue. For instance, a more effective review would explicitly recommend refactoring the conditional logic using Python's \texttt{or} operator, as demonstrated in the refined version (final line in expected refined code). This example highlights a critical limitation of descriptive feedback: it lacks the specificity needed to guide meaningful code refinement. Without actionable suggestions, automated review systems struggle to bridge the gap between identifying problems and enabling accurate code refinement.

\subsection{Example 2: Semantic Equivalence Challenge} 

This example highlights another critical issue in training automated code review systems: learning methods that prioritize exact matches between generated and real review comments, despite the possibility of semantically equivalent but differently phrased feedback. Such strategies can hinder the quality of produced models by penalizing valid variations in wording, ultimately limiting their ability to generate diverse yet functionally-equivalent reviews.

\begin{tcolorbox}[
    title=Example 2,
    colframe=black!30,
    colback=white,
    coltitle=black,
    fonttitle=\bfseries,
    sharp corners,
    boxrule=0.5pt,
    top=8pt,
    bottom=8pt
]
\textbf{Initial Code:}
\begin{lstlisting}[language=Python, basicstyle=\ttfamily\small]
 	arguments := make([][]byte, len(request.Args))
 	for i, arg := range request.Args {
-		argBytes, err := hex.DecodeString(arg)
+		var argBytes []byte
\end{lstlisting}

\textbf{Human review:} 
``can you move the declaration outside of the loop?''

\textbf{Potential generated review:} 
``Consider initializing the variable before the 'for' block to prevent redundant reassignments''

\textbf{Expected refined code:}
\begin{lstlisting}[language=Python, basicstyle=\ttfamily\small]
 	arguments := make([][]byte, len(request.Args))
+	var argBytes []byte
 	for i, arg := range request.Args {
-		var argBytes []byte
 		argBytes, err = hex.DecodeString(arg)
\end{lstlisting}
\end{tcolorbox}

\noindent\textbf{Challenge 2.} The human review and generated review, while phrased differently, convey identical semantic intent: moving the variable declaration outside the loop to improve efficiency. However, traditional learning methods and training strategies favor exact matches between generated and real comments, penalizing valid variations in wording. This narrow focus can hinder the quality of produced models, as it discourages the generation of semantically equivalent but differently phrased feedback. This limitation underscores the need for learning approaches that prioritize semantic similarity over syntactic match, enabling models to generalize and generate diverse yet accurate feedback that aligns with real-world developer practices.
\section{\oapp: Code Review Automation with reinforcement Learning}
\label{sec:framework}

Most deep learning models undergo training using their own feedback mechanism, wherein their predictions are juxtaposed with the actual output. However, alternative mechanisms leverage additional feedback sources to render the signal more informative than a mere comparison with the ground truth and thus improve the relevance of the predictions. These involve utilizing feedback signals from other models. The training process can take the form of collaboration, adversary, or reinforcement to achieve the target task.

In collaborative training, models cooperate through feedback to achieve better performance. For example, knowledge distillation is a collaborative learning technique that involves transferring knowledge from a large model (\ie teacher) to a smaller model (\ie student) \cite{hinton2015distilling}. The goal is to achieve comparable or even better performance on a target task using a smaller model. The process of knowledge distillation consists of training a teacher model on a large dataset to accomplish a specific task. The student model is then trained to mimic the behavior of the teacher model by minimizing the distance between their output distributions on the same dataset, typically using the Kullback-Leibler divergence (KL-divergence) as the distance metric \cite{joyce2011kullback, kim2021comparing}.

In competitive training, models compete to outperform each other.
For example, generative adversarial networks (GANs) are used for generative tasks to create new data that resemble the original data distribution \cite{goodfellow2020generative}. The typical GAN architecture consists of two neural networks: a generator and a discriminator. The generator learns to create synthetic data, while the discriminator learns to distinguish between the generated and the real data.
GANs are trained in an adversarial and competitive manner. The generator attempts to fool the discriminator. The discriminator aims to improve its ability to differentiate between real and fake data. Consequently, both models improve with training over time. 

RL presents another paradigm where models learn to make decisions by maximizing cumulative rewards. Unlike supervised learning, which relies on labeled data, RL involves training an agent to interact with an environment, taking actions that lead to desired outcomes. The agent receives feedback in the form of rewards or penalties based on its actions, which guides its learning process. This approach is particularly effective in scenarios where the optimal strategy involves a sequence of decisions, as the agent learns to anticipate long-term consequences and adjusts its actions accordingly. By iteratively refining its policy to maximize rewards, the RL agent becomes proficient in navigating complex environments and solving tasks that require strategic planning and adaptability. For example, reinforcement learning from human feedback is a method that uses RL to align trained models to human preferences \cite{ouyang2022training}. This method is promising and impactful, as it was employed to finetune ChatGPT on human preferences.

Software engineering tasks have traditionally been addressed in isolation, with each DL model either fine-tuned or trained for individual tasks. However, this overlooks the potential benefits of inter-task feedback and other feedback signals that can significantly enhance task performance. That is, integrating feedback mechanisms and using diverse signals from related tasks can lead to substantial improvements in the robustness and efficiency of related tasks \cite{sghaier2024improving}. This would produce models that are better equipped to capture complexities and interdependencies in software engineering tasks, thereby leading to more effective and intelligent automation.

In this section, we introduce a novel architecture, named \oapp, for review comment generation. We leverage different reward strategies for guiding this task.
First, we present the general architecture of \oapp.
Subsequently, we detail its different components and training phases.

\subsection{Framework overview}

\begin{figure*}[!htbp]
    \centering
    \includegraphics[width=1\textwidth]{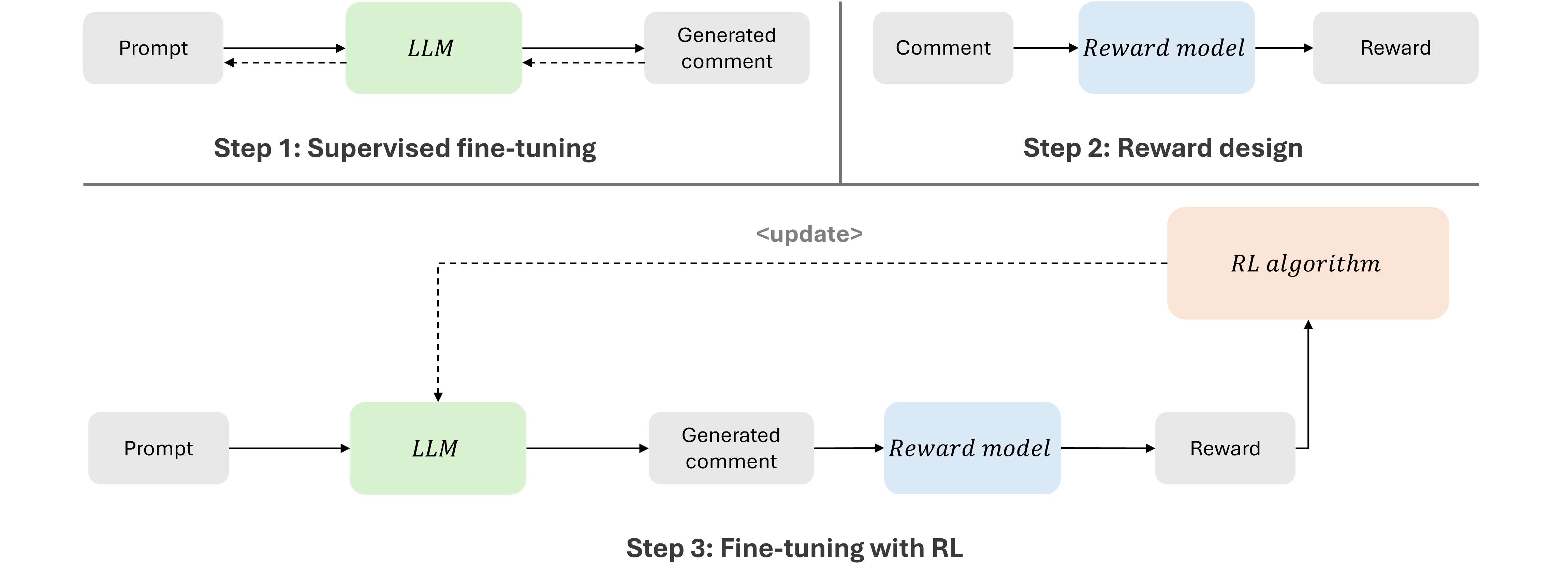}
    \caption{\oapp: framework overview}
    \label{fig:framework_overview}
\end{figure*}

We introduce \oapp, a RL-based framework for training LLMs for review comment generation. \oapp leverages different reward models to improve the efficiency of LLMs in this task.
\Fig{fig:framework_overview} shows an overview of our framework composed of three main steps.

The first step is a supervised fine-tuning of an LLM on the comment generation task. A prompt is fed to the LLM containing the code submitted for review, and the LLM generates a review comment. 
The second step is the reward design. It consists of creating a reward model or function that evaluates the comment according to specific criteria. The reward model generates a reward (\ie score) to reflect the relevance of the input comment with respect to the defined criteria.
The third step consists of a second fine-tuning phase of the LLM with RL and the designed reward model. A prompt containing the code is fed to the LLM, fine-tuned in the first step, to generate a review comment. The LLM weights are updated to maximize the reward metrics.

% Integrate this in the introduction
% Writing a loss function to capture these attributes seems intractable, and most language models are still trained with a simple next token prediction loss (e.g. cross entropy). To compensate for the shortcomings of the loss itself people define metrics that are designed to better capture human preferences such as BLEU or ROUGE. While being better suited than the loss function itself at measuring performance these metrics simply compare generated text to references with simple rules and are thus also limited. Wouldn't it be great if we use human feedback for generated text as a measure of performance or go even one step further and use that feedback as a loss to optimize the model? That's the idea of Reinforcement Learning from Human Feedback (RLHF); use methods from reinforcement learning to directly optimize a language model with human feedback. RLHF has enabled language models to begin to align a model trained on a general corpus of text data to that of complex human values.

\subsection{Supervised fine-tuning}

In the first step of our framework, we perform supervised finetuning of a large language model (LLM) specifically for the task of review comment generation. This involves training the LLM on a curated dataset that includes pairs of submitted code changes and their corresponding review comments. By exposing the model to a diverse array of examples, it learns to generate contextually appropriate and informative comments.

During the supervised fine-tuning phase, the LLM receives a prompt containing the code submitted for review. The model then processes this input and generates a corresponding review comment. This generated comment is compared to the ground truth comment from the dataset. The objective during this phase is to minimize the difference between the generated comment and the ground truth, typically using a loss function such as cross-entropy loss. This loss is then backpropagated through the model to update its weights, thereby improving its ability to generate accurate review comments over time.

Mathematically, let $\mathcal{D}$ denote our dataset composed of pairs $(\delta c, r)$ where $\delta c$ is the code change (\ie code difference) and $r$ the corresponding review comment.The LLM is finetuned by minimizing the cross-entropy loss function \( \mathcal{L} \), defined as:

\begin{equation}
    \mathcal{L}(\theta) = -\sum_{i=1}^{N} \sum_{t=1}^{K} \sum_{j=1}^{M} y_{itj} \log P_\theta(y_{itj} | \delta c_i)
\end{equation}

where:
\begin{equation*}
\left\{
    \begin{array}{ll}
         \theta & \text{represents the model parameters.} \\
         N & \text{is the number of training examples.}\\
         M & \text{is the vocabulary size.}\\
         K & \text{\parbox[t]{6cm}{is the maximum sequence length (\ie maximum review comment length).}}\\
         y_{itj} & \text{\parbox[t]{6cm}{is the ground truth indicator (1 if the $j^{th}$ word  is the correct word for the position $t$ in the $i^{th}$ review comment $r_i$, $0$ otherwise).}}\\
         P_\theta(y_{itj} | X_i) & \text{\parbox[t]{6cm}{is the probability assigned by the model to the word $y_{itj}$ given the input code $\delta c_i$.}}\\
    \end{array}
\right.
\end{equation*}

\smallskip

The cross-entropy loss measures how well the predicted probability distribution matches the actual distribution of the review comments, guiding the LLM to generate more accurate and contextually relevant comments.

\subsection{Reward model design}
\label{sec:reward_design}

\begin{figure*}[!htbp]
\centering
\begin{subfigure}{1\linewidth}
  \centering
  \includegraphics[width=1\textwidth]{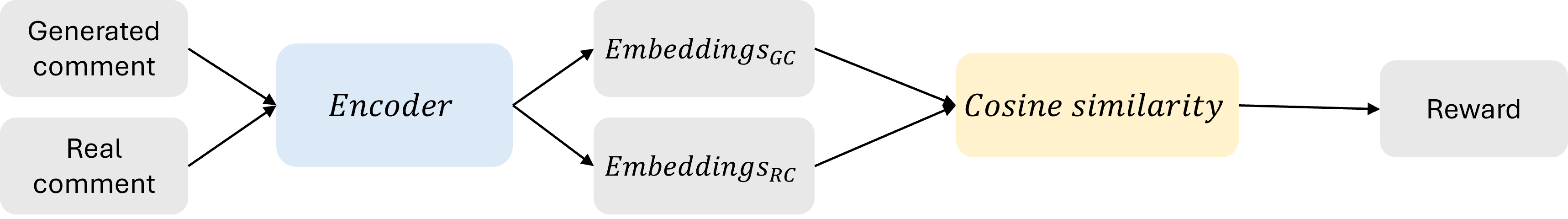}
  \caption{Reward strategy 1: semantic similarity}
  \label{fig:reward1}
\end{subfigure}
\vspace{10pt}
\begin{subfigure}{1\linewidth}
  \vspace{10pt}
  \centering
  \includegraphics[width=1\textwidth]{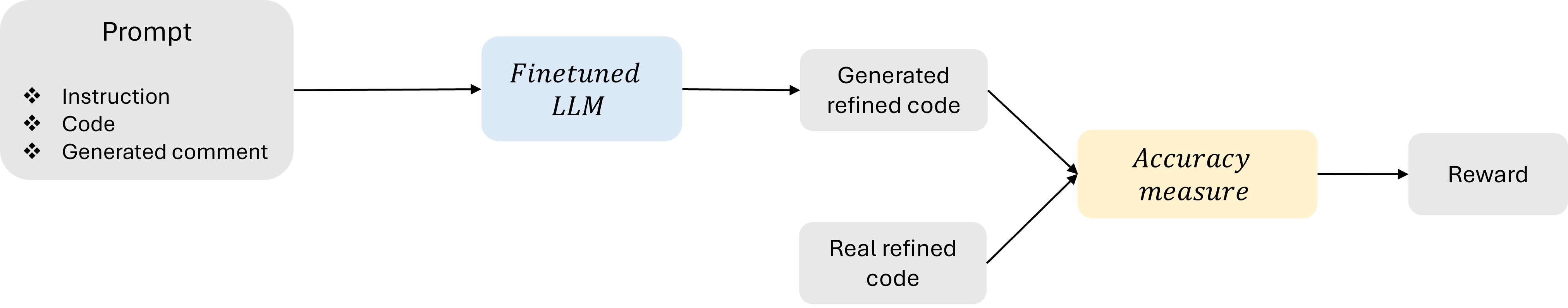}
  \caption{Reward strategy 2: correctness of the subsequent task}
  \label{fig:reward2}
\end{subfigure}
\caption{Reward models design}
\label{fig:rewards}
\end{figure*}

The loss function in LLMs is traditionally designed to evaluate the exact correspondence between predicted outputs and the ground truth. Nonetheless, a multitude of other characteristics contribute to the quality of predictions, specifically review comments. Semantic similarity, for example, is an important characteristic as it allows the predicted comment to diverge lexically from the ground truth while maintaining equivalent meaning through varied syntactic constructions. Additionally, the relevance of the review comment to subsequent tasks, such as code refinement, is critically significant.

%Writing a loss function to capture these attributes seems intractable. To compensate for the shortcomings of the loss itself, some metrics can be exploited, as documented in the next section.

The second step in our framework involves designing a reward model that will be used as feedback during the next RL phase. The reward model assesses the generated comments based on specific criteria such as relevance, clarity, and usefulness, assigning a score that reflects the overall quality of the comment.

\subsubsection{\textbf{Reward model strategies}}
We defined two distinct reward strategies to enhance the quality of review comments: \emph{similarity-based reward} and \emph{subsequent task correctness-based reward}.

\emph{Similarity-based reward} focuses on computing the semantic similarity between the predicted and actual review comments. The goal is to ensure that the generated comment conveys the same meaning as the ground truth, even if the wording differs. By prioritizing semantic similarity, the model can produce review comments that are flexible in their phrasing, yet consistent in their conveyed meaning.

The second strategy assesses the correctness of the subsequent task, \ie code refinement. Here, the predicted comment is used as input for the next task, and feedback is obtained on its usefulness and relevance. The goal of this approach is to ensure that the review comment not only reflects accurate information but also effectively guides the subsequent refinement process. We employ two methods to measure correctness: the loss value of the subsequent task and the CrystalBLEU score \cite{eghbali2022crystalbleu} (\ie an enhanced version of BLEU specifically designed to capture code similarity) between the real refined code and the predicted code edits. The loss value provides a quantitative measure of how well the refinement task is performed, while the CrystalBLEU score evaluates the closeness of the predicted edits to the actual refined code, thus ensuring practical applicability.

\subsubsection{\textbf{Reward models implementation}}
\label{sub:rewardStrategies}
\Fig{fig:reward1} illustrates the implementation details of the first reward model, \ie semantic similarity. First, we encode both the predicted and actual review comments using a sentence transformer (\ie SBERT). The generated embeddings are high-dimensional vectors that capture the semantic meaning of the comments. By computing the cosine similarity between these vectors, we derive a score that indicates how semantically close the predicted comment is to the actual comment. This score is then used as the reward signal during reinforcement learning, guiding the model to generate comments that are semantically similar to high-quality examples.

\Fig{fig:reward2} shows the implementation details of the second reward strategy, which assesses the correctness of the subsequent task. We integrate the predicted review comment into the code refinement task. We fine-tune an LLM on code refinement and use it as a reward model.
The LLM receives a prompt that includes the instruction, the initial version of the code, and the review comment. It then generates a refined version of the code, which is compared with the actual refined version using specific accuracy measures to determine the reward score. We employ two different accuracy measures: the loss value and CrystalBLEU.
The loss value of the subsequent task provides a direct measure of task execution quality with the given input. A lower loss value signifies better performance, leading to a higher reward. CrystalBLEU compares the predicted refined code with the actual one. A higher CrystalBLEU score indicates greater similarity to the ideal output, ensuring that the generated comments result in accurate and effective code refinement.

\subsection{Fine-tuning with reinforcement learning}

The third and final step in our framework involves a second phase of finetuning the LLM using RL with the designed reward model. In this phase, the LLM is further refined to maximize the rewards provided by the reward model, thereby enhancing its performance in generating review comments.

\begin{figure}[!htbp]
    \centering
    \includegraphics[width=1\linewidth]{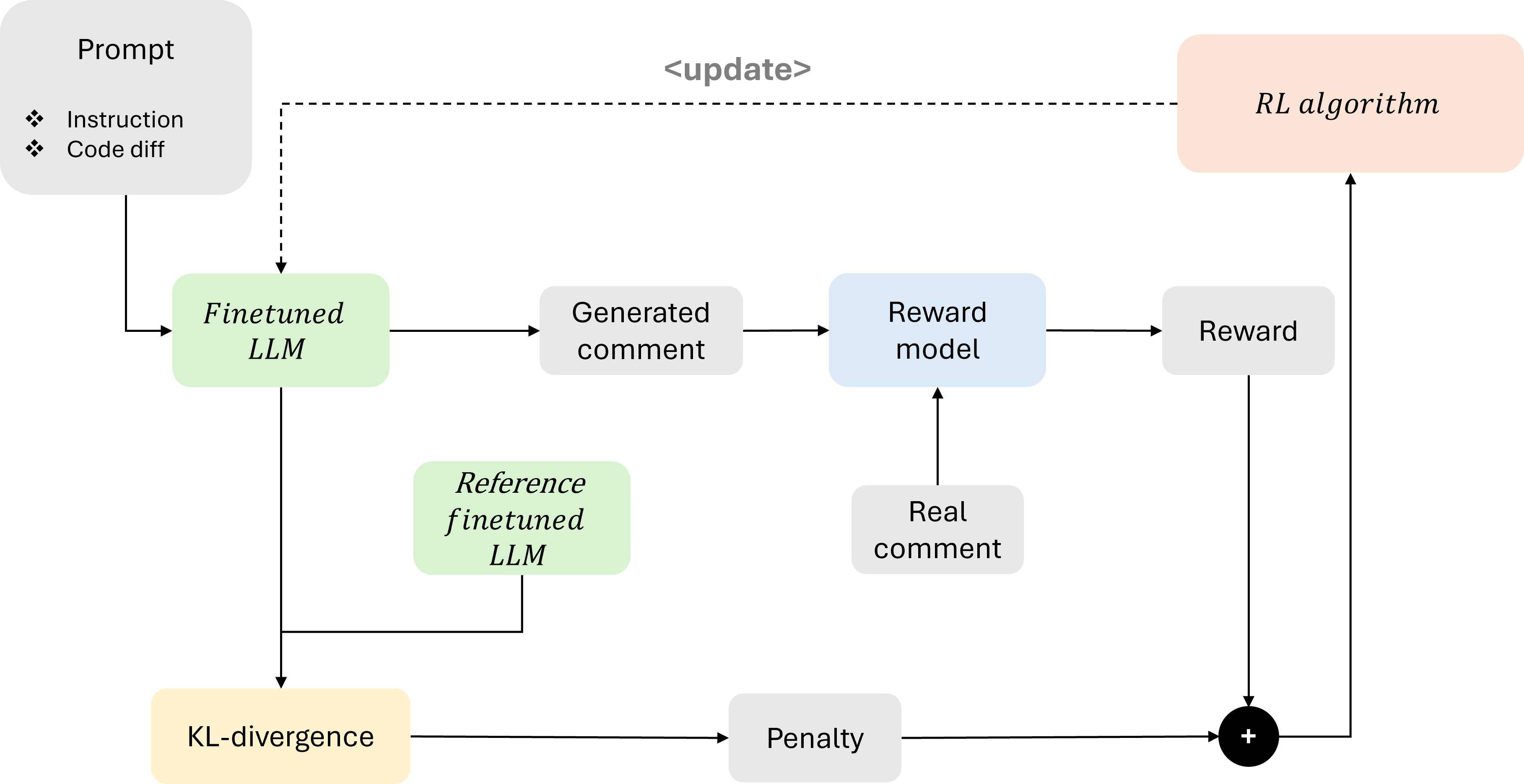}
    \caption{Detailed fine-tuning phase  with reinforcement learning}
    % \\The forward arrows represent the forward pass, the backward dashed arrows represent the backpropagation of the model....}
    \label{fig:framework_rl}
\end{figure}

As shown in \Fig{fig:framework_rl}, the RL finetuning process begins by feeding a prompt containing the instruction and the code difference to the LLM, which was already finetuned in the first step on comment generation. The LLM generates a review comment based on this input. This generated comment is then evaluated by the reward model, which assigns a reward score.

Similarly to \cite{ouyang2022training}, we incorporate a penalty mechanism to ensure stability and coherence in the model outputs. Specifically, the per-token probability distributions generated by the RL policy (\ie LLM) are compared with those from the initial model. The penalty is computed based on the scaled \emph{Kullback–Leibler (KL)} divergence between these sequences of distributions. This term discourages the RL policy from deviating significantly from the initial pretrained model during training \cite{gao2023scaling, glaese2022improving, bai2022training}, which helps maintain the generation of coherent text snippets. Without this penalty, the model might produce meaningless text that scores high with the reward model. %In practice, the KL divergence is approximated by sampling from both distributions. 
%This technique, validated in multiple studies , ensures that the model remains aligned with the pretrained behavior (\ie generates coherent text) while optimizing for the reward.

The final reward is the difference between the score generated with the designed reward model and the penalty score (\ie KL divergence).

\begin{equation}
    r = r_\theta - \lambda r_{KL}\\
\end{equation}

with $r_\theta$ being the score generated by the designed reward model in the second phase and $r_{KL}$ the penalty shift calculated with the KL divergence between the trained policy ($\Pi$) and the initial model ($\Pi_{base}$).

\begin{equation}
    r_\theta = \mathcal{R}(r)\\
\end{equation}
\begin{equation}
    r_{KL} = -D_{KL}(\Pi(y|x), \Pi_{base}(y|x))\\
\end{equation}

The objective of the LLM during this phase is to maximize the reward score through iterative updates to its weights using a RL algorithm (\eg proximal policy optimization \cite{schulman2017proximal}).

\section{Study design}
\label{sec:design}
The \emph{goal} of our study is to assess the effectiveness of \oapp in generating meaningful comments for a given code submitted for review. The \emph{context} consists of (i) a dataset of $176,616$ code review rounds from \cite{li2022automating}, where a review round corresponds to a triplet featuring the code submitted for review, a review comment, and a revised code implementing the comment; and (ii) DISCOREV, an approach recently proposed by Sghaier \etal \cite{sghaier2024improving}, being the current state of the art in code review comments generation.

Specifically, we formulate the following research questions (RQs):

\begin{itemize}
    \item \textbf{RQ$_1$}: \emph{Is the reinforcement learning fine-tuning boosting the performance of \oapp?} We perform an ablation study aimed at investigating whether the RL-based fine-tuning actually has a positive impact on the capabilities of the model in generating meaningful review comments.
    \item \textbf{RQ$_2$}: \emph{What is the best reward strategy to adopt in \oapp?} We compare the quality of the comments generated by \oapp with the two different reward strategies described in Section \ref{sec:reward_design}.
    \item \textbf{RQ$_3$}: \emph{How does \oapp compare to DISCOREV \cite{sghaier2024improving} in code review comments generation?} We compare the quality of the comments generated by \oapp with those produced by DISCOREV, the state-of-the-art approach for comment generation.
    \item \textbf{RQ$_4$}: \emph{How useful are the review comments generated by \oapp compared to the baseline?} We exploit as evaluation an LLM-as-judge approach, in which o3-mini has been prompted to assess whether the review comments generated by \oapp are more/less useful than those generated by the baseline.
   
\end{itemize}

\subsection{Context Selection: Dataset}
We rely on the same dataset used in the most recent works on code review automation \cite{li2022automating, sghaier2024improving}. The dataset has been originally presented in \cite{li2022automating}, and features $176,616$ code review rounds mined from public GitHub projects written in nine different languages. Each code review round can be represented as a triplet ($c$, $r$, $c_r$), with $c$ being the original code submitted for review, $r$ a review comment, and $c_r$ a revised version of $c$ aimed at addressing $r$ (\ie implementing the code changes required by the reviewer).

As done in previous work, we split the dataset into training ($85\%$), test ($7.5\%$), and validation ($7.5\%$). Adopting exactly the same split allows for simple comparison with the results previously reported in the literature for other code review comment generation techniques \cite{li2022automating, sghaier2024improving}.

\subsection{Context Selection: LLM}
As the LLM at the core of \oapp (see green boxes in Figure \ref{fig:framework_overview}), we adopt CodeLlama-7B \cite{roziere2023code}. CodeLlama is a family of LLMs for code built on top of Llama2 \cite{touvron2023llama}. CodeLlama has been trained on a corpus of $500B$ tokens featuring $85\%$ of code, $8\%$ of natural language related to code, and $7\%$ of natural language. The interested reader can find information about the architectural details of CodeLlama-7B (\eg number of layers, neurons, etc.) in \cite{roziere2023code}.

While larger variants of CodeLlama exist (up to $70B$), in \oapp we adopt a smaller version due to the high computational cost of fine-tuning these models. Indeed, just fine-tuning CodeLlama-7B for the task of comment generation (\ie excluding all fine-tuning via RL) took $113$ hours on a server equipped with four \emph{NVIDIA RTX A5000} graphics processing units (GPUs).

\subsection{CodeLlama Fine-tunings}
We train two different CodeLlama-7B models, one for comment generation and one for code refinement. For both models we used the following hyperparameter settings. We conducted training of CodeLlama using four \emph{NVIDIA RTX A5000} GPUs, with a batch size of $4$ per device. To enhance training efficiency, we employed several techniques. Gradient accumulation steps of $4$ were utilized, where gradients are accumulated over multiple batches, and the optimizer is updated only after a specific number of batches. We implemented 4-bit quantization to further improve memory efficiency and computational speed. Additionally, we employed Low-Rank Adaptation (LoRA) \cite{hu2021lora}, a Parameter-Efficient Fine-Tuning (PEFT) technique, configured with settings of $r=16$, $\alpha=32$, and $dropout=0.05$. LoRA operates by decomposing the weight updates of a neural network into low-rank matrices, significantly reducing the number of parameters that require updating during fine-tuning \cite{hu2021lora}, thus enhancing the overall efficiency of the training process.

When training for code refinement, we provide the model with a pair ($c$, $r$) as input (\ie the code submitted for review and a reviewer comment) and ask the model to generate $c_r$ (\ie the refined version of the code addressing $r$). Note that $c$ includes both the complete source code file on which the comment $r$ has been posted as well as a \texttt{diff} showing the code changes. We trained the model for five epochs on the previously described training set. Early stopping was enabled to prevent overfitting, ensuring the model stopped training once performance on the validation set, as measured by the loss, ceased improving. We use an early stopping patience of $20$, to stop training when the specified metric worsens for $20$ steps.
This model, trained for code refinement, will be used in the context of the RL finetuning to provide rewards for the comments generated by \oapp, as shown in Figure~\ref{fig:reward2}. 

As for the comment generation task, being the focus of our approach, we start by training CodeLlama for the task of review comment generation using the standard fine-tuning procedure also adopted in previous work \cite{sghaier2024improving, tufano2022using, li2022automating}: The model is provided with the code change $c$ (\texttt{diff}) submitted for review as input and it required to generate $r$. 
We opt for a similar training procedure as for code refinement, \ie we trained the model for five epochs with early stopping enabled.
This model represents both the starting point for the further RL-based fine-tuning described in the following as well as a baseline through which we can properly assess the contribution given by the RL step.

\subsubsection{Fine-tuning via RL}
As described in Section \ref{sec:reward_design}, we experiment with two different reward strategies. The first is based on the semantic similarity between the comments generated by the fine-tuned model and the target comment, as assessed via SBERT \cite{reimers2019sentence}. The second, instead, exploits the CodeLlama model fine-tuned for the code refinement task as detailed in \Sect{sub:rewardStrategies} (see the two variants using the loss function and the CrystalBLEU score as reward). 

To perform RL-based fine-tuning, we used the Transformer Reinforcement Learning (TRL) library of huggingface. We trained CodeLlama using three \emph{NVIDIA RTX A5000} GPUs with a batch size of $4$ per device. The fourth GPU was dedicated to the reward model.
To improve training efficiency, we used the same optimizations previously described for the standard fine-tuning (\ie gradient accumulation, bit quantization, LoRa). The Proximal Policy Optimization (PPO) algorithm \cite{schulman2017proximal} was used for training, ensuring robust policy optimization. During training, responses (\ie review comments) were generated, rewards were computed, and the model was updated iteratively based on these rewards. Additionally, we used a learning rate of  \(5e-5\), AdamW optimizer, and gradient checkpointing to manage memory usage during backpropagation. Early stopping was implemented to prevent overfitting, and periodic checkpointing was used to save the model at regular intervals.

\subsection{Data Collection and Analysis}
We answer RQ$_1$ and RQ$_2$ by comparing the BLEU score \cite{papineni2002bleu} (between the generated and the target comments) achieved by the CodeLlama model fine-tuned for the comment generation task \emph{without} the further RL-based fine-tuning, and three variants of our RL-based approach, namely those exploiting the rewards provided via (i) SBERT semantic similarity, and (ii) the loss and the (iii) CrystalBLEU of the subsequent code refinement task. This allows to see if the RL-based fine-tuning had any positive impact on the performance of the model (RQ$_1$) as well as which of the reward models we experimented with is the best one (RQ$_2$).

We report boxplots showing the BLEU distributions the four above-described models achieve on the test set. We also statistically compare these distributions using the Mann-Whitney test \cite{fagerland2009wilcoxon}. We account for multiple tests by adjusting \emph{p}-values using the Benjamini-Hochberg procedure \cite{benjamini1995controlling}. We use the Cliff's $d$elta \cite{macbeth2011cliff} as effect size. Cliff's $d$ ranges in  the interval $[-1,1]$ and is negligible for $|d| < 0.148$, small for $0.148 \le |d| < 0.33$, medium for $0.33 \le |d| < 0.474$, and large for $|d| \ge 0.474$.

As for RQ$_3$, we compare \oapp (in its best setting as identified in RQ$_2$) with the state-of-the-art technique recently proposed by Sghaier \etal \cite{sghaier2024improving} in terms of BLEU score. We also use some statistical tests (\ie Mann-Whitney test \cite{fagerland2009wilcoxon} and Cliff's $d$elta \cite{macbeth2011cliff}) to compare the distributions of the results for both models.

Besides the quantitative analysis done in RQ$_1$ and RQ$_2$, we also perform a more ``qualitative'' evaluation in RQ$_4$, asking Open AI o3-mini model to act as judge and assess the usefulness of the comments generated by our framework \oapp~and the baseline (CodeLlama model finetuned for the comment generation task). o3-mini judges which generated comment is more useful or if they are equally useful.

In this experiment, we assume that a highly capable model, specifically o3-mini, can serve as a substitute for human evaluators to accurately assess the relevance of generated comments. This reliance on o3-mini is supported by previous research demonstrating that GPT3.5 and GPT4 closely align with human judgments \cite{zheng2024judging, li2023alpacaeval}, achieving agreement rates comparable to human-to-human agreement on MT-Bench \cite{zheng2024judging}. Such findings justify their use as evaluators in our assessments. In particular, GPT4 has been recently exploited as judge for other software engineering tasks, such as in\cite{weyssow2024codeultrafeedback}. The authors used it to assess the extent to which an automatically implemented code meets specific non-functional requirements (\eg comprehensibility) \cite{weyssow2024codeultrafeedback}. 

To further ensure the reliability of o3-mini assessments, we performed a sanity check by manually assessing a random sample of 100 pairs of comments, deciding which comment was more useful. We then evaluated the alignment of o3-mini decisions with human judgments using \emph{Cohen's kappa}, a statistical measure used to evaluate the level of agreement between two raters, accounting for the possibility of the agreement occuring by chance. By calculating \emph{Cohen's kappa}, we can qualitatively assess the consistency and performance of o3-mini compared to human evaluators, thereby providing a robust validation of our automated assessment methodology.

We used the following prompt to trigger the judgment task:

\begin{table}[h]
  \centering
  \caption{Prompt template used for the judgment task}
  \label{tab:dataset}
  \begin{tabularx}{\linewidth}{X}
  \toprule
    Code review is a software quality assurance activity in which one or more developers (named reviewers) inspect the code changes implemented by a teammate (named contributor) with the goal of identifying code quality issues and suboptimal implementation choices.
    
    \vspace{3pt}
    
    When reviewers spot an issue, they write a natural language comment describing the issue and, possibly, suggesting how to address it. The comment is then used by the contributor to revise the code, addressing the highlighted issue.

    \vspace{3pt}

    A reviewer's comment can be considered useful if it identifies a quality issue relevant to the implemented change, it clearly and succinctly describes it, and results in an actionable suggestion for the contributor.

    \vspace{5pt}

    Given the following Java file: 
    $<$start\_code$>$\{code\}$<$end\_code$>$

    On which the following change has been performed: $<$start\_diff$>$\{diff\}$<$end\_diff$>$

    \vspace{5pt}
    
    Can you assess which of the following two reviewer's comments is more useful?
        
        Comment 1: \{comment1\}
        
        Comment 2: \{comment2\}

    Answer by outputting 1, if the first comment is more useful, 2 if the second comment is more useful, 0 if they are considered equally useful.
    \\\bottomrule
  \end{tabularx}
\end{table}

This judgment task was run for all $13,104$ code reviews part of our test set. The order in which the two comments were presented was randomized, to avoid any sorting bias during the judgment. We answer RQ$_4$ by reporting the percentage of win, tie, and lose achieved by \oapp against the baseline.

\section{Results}
\label{sec:results}
\subsection{Results Discussion for RQ$_1$ and RQ$_2$}
\begin{figure}[!htbp]
    \centering
    \includegraphics[width=1\linewidth]{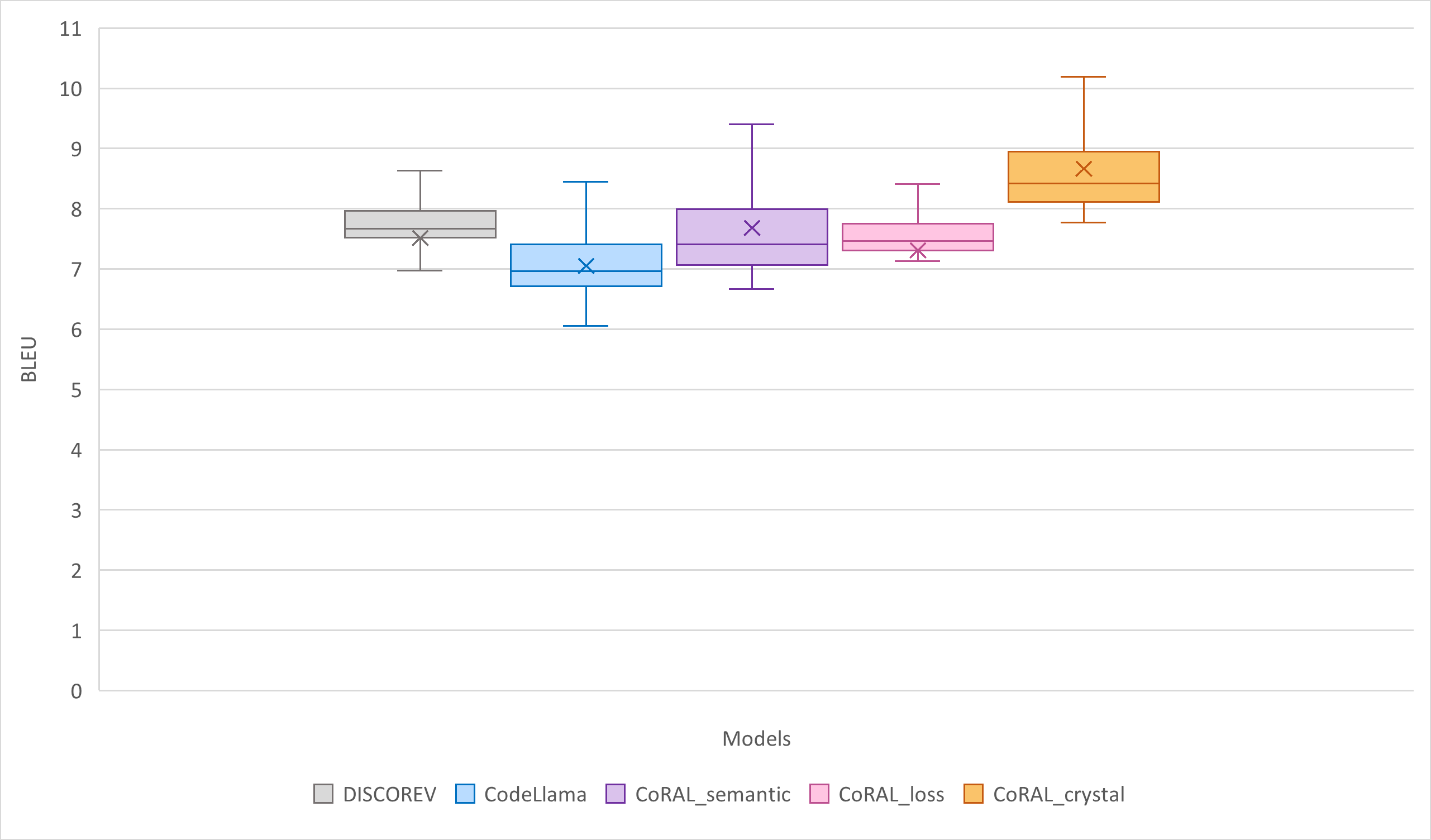}
    \caption{BLEU score distribution across the test set for the different models}
    % \\The forward arrows represent the forward pass, the backward dashed arrows represent the backpropagation of the model....}
    \label{fig:results1}
\end{figure}

\Fig{fig:results1} illustrates the distribution of BLEU scores on the test set for the different models.

% \begin{figure*}[!htbp]
%     \centering
%     \includegraphics[width=1\linewidth]{figures/results1-v2.png}
%     \caption{BLEU score achieved by the different models for the comment generation task}
%     % \\The forward arrows represent the forward pass, the backward dashed arrows represent the backpropagation of the model....}
%     \label{fig:results11}
%     \obs{Should we include the SOTA works in the comparison to show the evolution since they are compared on the same test set? or should we just omit them? Also, should we just replace the DISCOREV trained using CodeT5 with DISCOREV trained using CodeLlama-7B or keep both?}
% \end{figure*}

\emph{CodeLlama\_sft}, our baseline, is the CodeLlama-7B fine-tuned for comment generation without RL. 
\emph{\oapp{}\_semantic} represents our RL-based approach utilizing \emph{semantic similarity} as reward, while \emph{\oapp{}\_loss} and \emph{\oapp{}\_crystal} represent our RL-based models trained using the \emph{loss} and the \emph{CrystalBLEU}, respectively, of the subsequent task (\ie code refinement) as a reward. 

As it can be seen, RL always helps to boost the performance of the model, independently from the specific reward function adopted. This indicates that the provided rewards represent valuable signals for the model, allowing it to improve its performance in comments generation. 

\emph{\oapp{}\_crystal} achieves the highest median BLEU score of $8.67$, outperforming \emph{\oapp{}\_semantic} at $7.68$, \emph{\oapp{}\_loss} at $7.31$, and \emph{CodeLlama\_sft} at $7.05$. This indicates that utilizing CrystalBLEU as the reward model provides a more effective feedback signal for generating high-quality comments. In contrast, the loss-based reward model proves to be less effective, likely due to the difficulty in interpreting the loss value meaningfully in the context of comment generation.

To validate our findings, we performed statistical tests as shown in \Table{tab:statsRQ1}. \emph{\oapp{}\_crystal} demonstrated a significant improvement over \emph{CodeLlama\_sft} with an adjusted $p\_value<0.001$ and a Cliff's delta of $-0.851$, indicating a substantial difference in BLEU scores. \emph{\oapp{}\_crystal} also significantly outperformed \emph{\oapp{}\_semantic} ($p\_value<0.001$, Cliff's delta $=-0.676$) and \emph{\oapp{}\_loss} ($p\_value<0.001$, Cliff's delta $=-0.840$).

% Although \emph{\oapp{}\_semantic} and \emph{\oapp{}\_loss} both outperformed \emph{CodeLlama\_sft}, their improvements were less pronounced (Cliff's delta $=-0.441$ and $-0.503$, respectively). The comparison between \emph{\oapp{}\_semantic} and \emph{\oapp{}\_loss} showed no significant difference ($p\_value=0.941$, Cliff's delta $=-0.091$), indicating similar performance between these two strategies, though both were outperformed by \emph{\oapp{}\_crystal}.

In conclusion, these statistical results affirm the superior performance of \oapp~compared to the baseline (\ie \emph{CodeLlama\_sft}), highlighting its effectiveness in generating more accurate and useful review comments.

\begin{table}[!htbp]
  \centering
  \caption{RQ$_1$ \& RQ$_2$: Mann-Whitney test (adj. p-value) and Cliff's Delta}
  \label{tab:statsRQ1}
    \begin{tabular}{l r r}
    \toprule
    \textbf{Test} & \textbf{Adj. $p$-value} & \textbf{Cliff's $d$}\\
    \midrule
    \oapp{}\_semantic \emph{vs} \emph{CodeLlama\_sft}& $<$0.001 & -0.441  \\
    \oapp{}\_loss \emph{vs} \emph{CodeLlama\_sft}    & $<$0.001 & -0.503 \\
    \oapp{}\_crystal \emph{vs} \emph{CodeLlama\_sft} & $<$0.001 & -0.851 \\\midrule
    \oapp{}\_semantic \emph{vs} \oapp{}\_loss    & 0.941    & -0.091\\
    \oapp{}\_crystal \emph{vs} \oapp{}\_semantic & $<$0.001 & -0.676\\
    \oapp{}\_crystal \emph{vs} \oapp{}\_loss     & $<$0.001 & -0.840\\
    \bottomrule
\end{tabular}
\end{table}

The improved effectiveness of the model in executing the required task is also demonstrated by the increase in the reward provided to it during the RL-based training. \Table{tab:results1} shows the initial and final reward provided to the \emph{CodeLlama} model, with the initial one being the average reward (across all test set instances) assigned to \emph{CodeLlama\_sft} (\ie the \emph{CodeLlama} model which underwent standard fine-tuning without any RL step) and the final one being the same metric computed after the RL-based training. 

\begin{table}[!htbp]
  \centering
\caption{Evolution of the rewards for the different models}
  \label{tab:results1}
    \begin{tabular}{l r r}
    \toprule
    \textbf{Model} & \textbf{Initial reward} & \textbf{Final reward}\\
    \midrule
    \oapp{}\_semantic & 0.18 & 0.29 \\
    \oapp{}\_loss & 0.69 & 0.68 \\
    \oapp{}\_crystal & 0.77 & 0.84 \\
    \bottomrule
\end{tabular}
\end{table}

As it can be seen, for all reward functions we observe an improvement. Indeed, for both semantic similarity and CystalBLEU higher scores are proxies for better predictions, while the opposite holds for the loss (\ie the lower the better).

For \emph{\oapp{}\_semantic}, we observe a notable increase in the average semantic similarity between the review comments generated by the model and the target (expected) ones in the test set. The increase goes from $0.18$ to $0.29$ (namely, a relative +61\%). This suggests that the model effectively learns to generate comments that are more semantically similar to the ground truth. 
When looking at \emph{\oapp{}\_loss}, the drop in loss indicates that the model learned to produce comments which allow a loss reduction of the subsequent task (\ie the generated comments are ``easier to implement'' for the model fine-tuned for the task of code refinement). While the change here may look small, even minor changes in the loss value may reflect in significant improvements when it comes to generative tasks. This is also confirmed by  \emph{\oapp{}\_crystal}, showing that the RL reward allows the code refinement model to better implement the generated comments, with an increase of the average \emph{CrystalBLEU} score from $0.77$ to $0.84$, indicating that the comments generated are more comprehensible and useful for the code refinement model.

In summary, we can positively answer \textbf{RQ$_1$} by observing that the RL-based fine-tuning helps in boosting the performance of DL-based review generation.

The above-discussed data also help in answering RQ$_2$. Indeed, \Fig{fig:results1} shows a clear trend in the BLEU score achieved by the experimented models, with \emph{\oapp{}\_crystal} being the best in class. This finding is also echoed by the results of the statistical analysis (\Table{tab:statsRQ1}) showing significant $p\_values$ not only when comparing \emph{\oapp{}\_crystal} to the baseline approach (\emph{CodeLlama\_sft}), but also when contrasting it against the other RL-based alternatives we experimented (\ie \oapp{}\_semantic and \oapp{}\_loss). The Cliff's delta of these comparisons ranges between -0.091 (\oapp{}\_semantic \emph{vs} \oapp{}\_loss) and -0.840 (\oapp{}\_crystal \emph{vs} \oapp{}\_loss).

For these reasons, we can answer \textbf{RQ$_2$} by stating that the reward strategy exploiting the CrystalBLEU of the subsequent (code refinement) task is the best one we experimented with and, as a consequence, will be the one we will consider when we qualitatively compare \oapp~against the baseline (\ie RQ$_4$).

\subsection{Results Discussion for RQ$_3$}

\emph{DISCOREV} \cite{sghaier2024improving} was originally trained with CodeT5. To ensure a fair comparison with \oapp, we re-trained \emph{DISCOREV} using \emph{CodeLlama-7B}, aligning the models  capacities.

As depicted in \Fig{fig:results1}, \emph{\oapp{}\_crystal} outperforms \emph{DISCOREV} in comment generation, evidenced by higher BLEU scores. \emph{\oapp{}\_crystal} achieves a median BLEU of $8.67$ compared to $7.51$ for \emph{DISCOREV}. This improvement is statistically significant, with $p\_value < 0.001$ and Cliff's delta $= 0.76$, indicating a large effect size. 

The substantial difference in BLEU scores indicates that \emph{\oapp{}\_crystal} model, which leverages CrystalBLEU, provides more effective feedback for generating high-quality comments. The high Cliff's delta value further underscores the large effect size, suggesting that \emph{\oapp{}\_crystal} consistently produces more relevant and accurate comments compared to \emph{DISCOREV}. This highlights the effectiveness of using CrystalBLEU as a reward model in reinforcement learning for comment generation. 

In conclusion, the results demonstrate that \emph{\oapp{}\_crystal} enhances the quality of generated comments compared to DISCOREV, the state-of-the-art model in comment generation.

\subsection{Results Discussion for RQ$_4$}

We perform a sanity check to validate the reliability of o3-mini in assessing the usefulness of comments. We randomly selected 100 pairs of review comments generated by \emph{\oapp{}\_crystal} and \emph{CodeLlama\_sft}. We manually judged them using a scale: 1 indicates that comment 1 is more useful, 2 indicates that comment 2 is more useful, and 0 indicates a tie.

\begin{figure}[!htbp]
    \centering
    \includegraphics[width=0.9\linewidth]{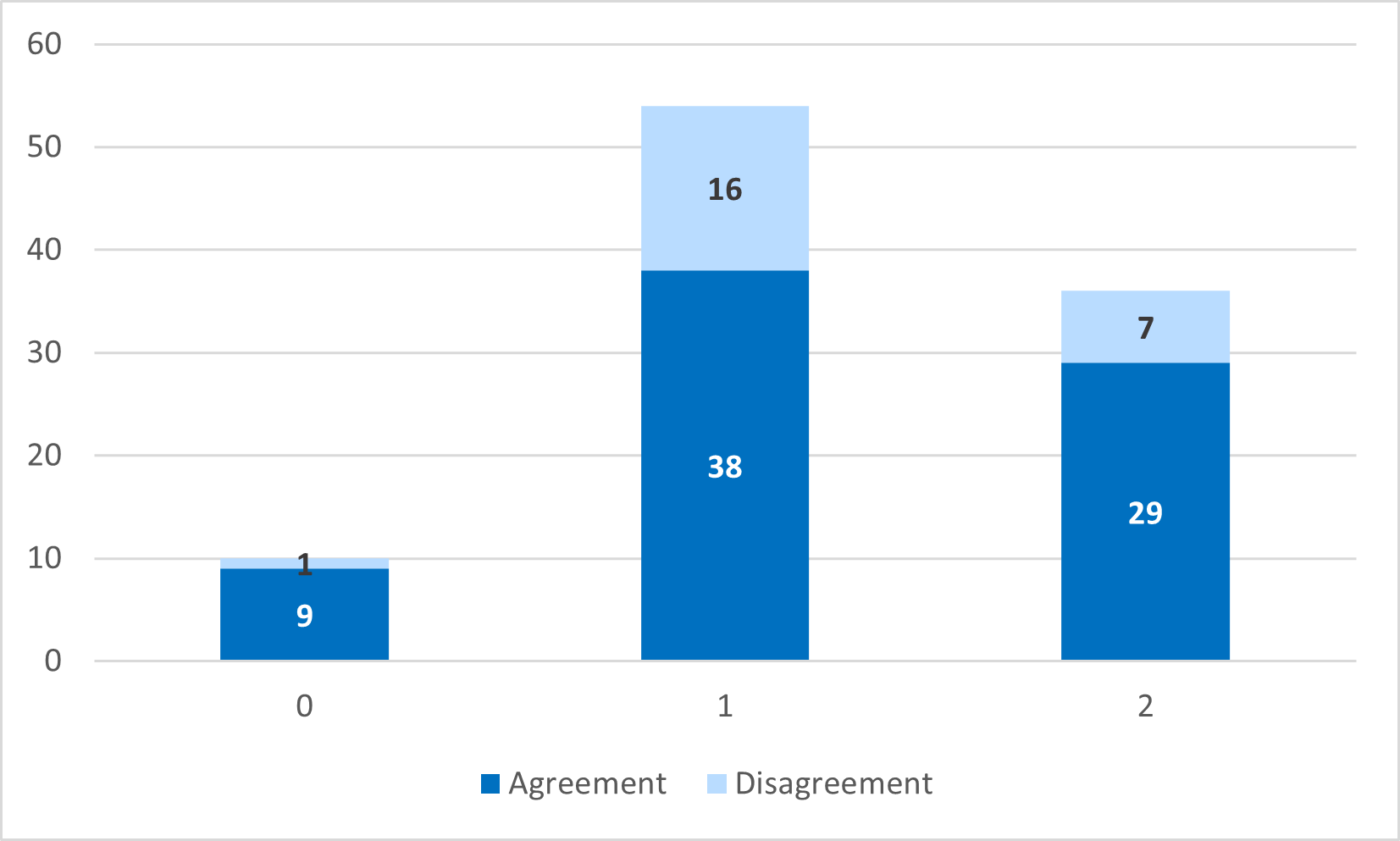}
    \caption{Agreement rates between human and o3-mini.}
    \label{fig:agreement}
\end{figure}

As shown in \Fig{fig:agreement}, we have agreement in $76\%$ of the cases. That is the human reviewer responded with the same judgment, as o3-mini, for $76\%$. 
Additionally, we calculated \emph{Cohen's kappa} coefficient. The resulting kappa value was $0.62$, which indicates a \emph{substantial level of agreement} according to the commonly accepted interpretation scale \cite{landis1977measurement}. This suggests that there is a substantial alignment between human and o3-mini judgments regarding the usefulness of the comments. This validates the reliability of o3-mini as a judge in this context and supports the findings in the literature \cite{zheng2024judging, li2023alpacaeval}.

\begin{figure}[!htbp]
    \centering
    \includegraphics[width=1\linewidth]{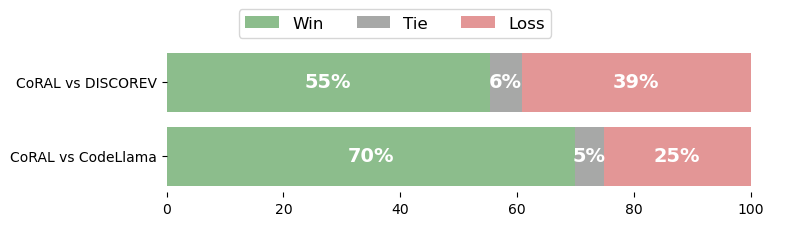}
    \caption{Results of OpenAI o3-mini judgments. A Win indicates that o3-mini preferred \oapp~comment; a Loss means the baseline comment was preferred; and a Tie indicates equal preference.}
    \label{fig:results2}
\end{figure}

\Fig{fig:results2} presents the results of \emph{o3-mini} judgments comparing \oapp~with two systems: CodeLlama-7B, used as a baseline, and DISCOREV \cite{sghaier2024improving}, a recent state-of-the-art approach for automated code review comment generation.
Each comparison was conducted on a set of 1000 comment pairs.
\oapp~was preferred in $70\%$ of the comparisons against CodeLlama-7B, while CodeLlama-7B was favored in only $25\%$ of the cases, with $5\%$ ties. Similarly, CoRAL outperformed DISCOREV in $55\%$ of the examples, lost in $39\%$, and tied in $6\%$. These results demonstrate the effectiveness of CoRAL in generating more relevant review comments than these baselines.

% \Fig{fig:results2} show the results of \emph{o3-mini} judgments on $1000$ pairs of comments generated by \emph{GPT4-turbo} and \emph{Codellama-7b}.
% \oapp generated more relevant comments in $516$ of the cases, as judged by \emph{GPT4-turbo}, compared to the baseline (\ie \emph{Codellama-7b}) that generated more useful comments for $399$ examples. The two models are tied, \ie, they generated equally useful comments, $77$ times. Note that \emph{GPT4-turbo} did not give a judgment for 8 instances. This proves the usefulness of our proposed framework in generating more relevant review comments than the baseline.

% \begin{figure}[!htbp]
%     \centering
%     \includegraphics[width=0.9\linewidth]{figures/gpt4_judgments-v2.png}
%     \caption{Results of GPT4-Turbo judgments.}
%     \label{fig:results2}
% \end{figure}

\section{Threats to validity}
\label{sec:threats}

The evaluation results have shown that our proposed framework is effective in addressing the comment generation task. However, there are certain threats that may limit the validity of these evaluation results.

A primary threat pertains to the nature of the data, specifically the reviews, which may contain noise and potentially non-English or misspelled words. We have mitigated this by employing \emph{CodeLlama-7B}, a state-of-the-art large language model that utilizes Byte-Pair Encoding \cite{sennrich2015neural}, a subword-based tokenization algorithm. This algorithm breaks unseen words into several frequently seen sub-words that can be effectively processed by the model.

A second concern revolves around the data imbalance, where some programming languages have fewer examples than others. This disparity may lead to varying performance across programming languages. However, the use of a large pre-trained language model allows us to circumvent this issue, as \emph{CodeLlama-7B} is a large language model that has already been trained on extensive code repositories. Consequently, fine-tuning this model for downstream tasks does not require a large volume of data. Moreover, prior research has shown that the use of multilingual training datasets can result in enhanced model performance compared to monolingual datasets, particularly for low-resource languages, in tasks such as neural machine translation and code translation \cite{chiang2021breaking, zhu2022multilingual}.

A third concern pertains to the selection of hyperparameters, which play a critical role in determining the model's performance. To ensure a fair comparison with \cite{sghaier2024improving}, we conducted a grid search solely on some hyper-parameters (\ie batch size, learning rate, gradient accumulation steps). It is important to note that further improvements may be achievable through additional hyperparameter tuning.

A final threat involves the size difference between our proposed framework \oapp, which uses \emph{CodeLlama-7B} (7 billion parameters), and DISCOREV \cite{sghaier2024improving}, which uses \emph{CodeT5} (220 million parameters). To ensure a fair comparison, we retrained DISCOREV with \emph{CodeLlama-7B}. This allows us to accurately compare the efficiency of both architectures by  using the same model.

\section{Related work}
\label{sec:literature}
\subsection{Automating Code Review}

To support developers in code review activities, researchers proposed techniques and tools to automate a variety of code review tasks. For example, several studies focus the attention on the reviewer recommendation task \cite{balachandran2013reducing, thongtanunam2015should, chouchen2021whoreview, Lee:fse2017, Al-Zubaidi:2020, li2023code}, namely the automatic selection of proper reviewers for a given code change. Others, instead, target the reviewer's comments classification task \cite{li2017automatic}, having the goal of automatically classifying the comments posted by reviewers based on the ``type of feedback'' they provide to the contributor (\eg feedback about the code style, functionality, etc.). 

Recently, research on code review automation started shifting from classification-based problems (like the ones discussed above), to more challenging generative tasks requiring the generation of textual content, such as review comments. The approaches in this context are mainly based on information retrieval (IR) and deep learning (DL). 

IR-based approaches assume that, in the context of code review, analogous situations may be encountered over time with, for example, similar code changes submitted for review. Representative of this line of works is the CommentFinder approach by Hong \etal \cite{hong2022commentfinder}, which retrieves reviewers' comments from the past that can be relevant for new code changes to review. Gupta and Sundaresan \cite{gupta2018intelligent} introduced DeepCodeReviewer (DCR), an LSTM-based model that is trained using positive and negative examples of (code, review) pairs. Given a new code snippet, a subset of candidate reviews is selected, from a predefined set of reviews, based on code similarity. Subsequently, DCR predicts a relevance score for each review (based on the input code snippet) and recommends reviews exhibiting high relevance scores. Siow \etal \cite{siow2020core} introduced a more sophisticated approach based on multi-level embedding to learn the relevancy between code and reviews. This approach uses word-level and character-level embeddings to achieve a better representation of the semantics provided by code and reviews. Clearly, the disadvantage behind IR-based techniques is that they cannot generate relevant comments for previously unseen code changes.

More relevant to our work are the techniques relying on DL-based techniques for generating review comments. Tufano \etal \cite{tufan2021towards, tufano2022using} presented an approach based on T5, a text-to-text transfer transformer which has been pre-trained with the masked language modeling task (\ie randomly masking 15\% of the works in sentences) to acquire general knowledge of the two languages of interest, namely Java (\ie the language used for the code to review) and English (\ie the language in which review comments must be generated). Along the same lines, Li et al \cite{li2022automating} pre-trained CodeT5 on four tasks, tailored specifically for the code review scenario, using a large-scale multilingual dataset of code reviews. Then, the output model was fine-tuned on three downstream tasks: quality estimation (\ie accept/reject a pull request), review generation (\ie generate review comment), and code refinement (\ie recommend code edits to satisfy the reviewer).
% Both these works are considered as baselines in our experiment.

More recently, other approaches have been proposed to improve performance in automating the same tasks \cite{sghaier2024improving}. In particular, Sghaier \etal \cite{sghaier2024improving} observed that quality estimation, comment generation, and code refinement are interconnected tasks and, thus, proposed DISCOREV an approach employing cross-task knowledge distillation to address them simultaneously. They utilize a cascade of models in which the fine-tuning of the comment generation model is guided by the code refinement model, while the fine-tuning of the code refinement model is guided by the quality estimation model. They show that DISCOREV improves the approaches by Tufano \etal \cite{tufan2021towards, tufano2022using} and Li \etal \cite{li2022automating}.

Our approach, while inspired by DISCOREV, introduces further enhancements by leveraging more sophisticated and diversified feedback signals beyond a simple combination of the loss functions. Our proposed framework, \oapp, integrates reinforcement learning to enhance comment generation through the optimization of rewards. We explore various reward mechanisms, \ie semantic similarity and correctness of the subsequent task.

\subsection{Reinforcement Learning to Automate Software Engineering Tasks}

RL-based agents have been shown to be particularly suited to learn how to play games, as shown by the impressive (sometimes superhuman) results reported in the literature \cite{Mnih2013, Mnih2015, Hessel2018, Baker:2020, Vinyals2017, berner:2019}. Based on these findings, software engineering researchers started using RL not only to play games but also to test them and, in general, to improve their quality. The basic idea is to automate playtesting using RL-based agents which can play the game as humans would do, thus helping in identifying possible quality issues \cite{Ariyurek:2019, Bergdahl:cog20, zheng:ase2019, wu:icsme2020, tufano:icse22}. For example, Zheng \etal \cite{zheng:ase2019} used evolutionary algorithms and multi-objective optimization to maximize the game exploration of a RL-based agent trained to play the game. During the training, heuristics are used to identify functional bugs spotted by the agent (\eg the game crashes). While the identification of functional bugs is the main aim pursued by this line of research, other works also focused on the identification of non-functional quality issues, such as performance bugs \cite{tufano:icse22}.

In addition to game testing, RL has also been exploited to support testing of other types of software \cite{aboeleneen:ist23}, mostly those requiring a GUI-based interaction \cite{harries:drift2020, collins:deep2021, mariani:2012, vuong:2018, adamo:2018, carino:2015} or being robotics-related, such as those behind autonomous vehicles \cite{feng2023dense, cho2020towards, koren2018adaptive}. Test case prioritization has also been addressed via RL \cite{bagherzadeh:tse22}.

Other applications of RL in software engineering pertain with code search \cite{ahmed:2023, sukur:sc2024}, software project management \cite{tlili:2021, chen:2024}, refactoring \cite{ahmadi:2022, haouari:2023} and models repair \cite{barriga:models2020}. More relevant to our work are RL-based techniques aimed at automatically generating textual content, such as those proposed for code summarization \cite{wan:ase2018, wang:tse22} and code generation \cite{le2022coderl, wang:cc2022}. For example, Le \etal \cite{le2022coderl}, in the context of code generation, proposed an approach featuring collaboration between language models and RL. In particular, during the training phase, a critic network is used to predict the functional correctness of the generated programs thus providing feedback to the language model generating the code. During the inference phase, instead, the model automatically regenerates programs based on the received feedback.

While related to our work, to the best of our knowledge \oapp is the first RL-based technique proposed in the literature for the task of review comment generation.
\section{Conclusion}
\label{sec:conclusion}

In this paper, we introduce \oapp, a novel framework using reinforcement learning to improve the generation of code review comments. Our framework employs two distinct reward strategies: semantic similarity and subsequent task. The semantic similarity strategy aims to ensure that generated comments closely resemble real reviews in terms of meaning. The subsequent task strategy emphasizes the accuracy of subsequent code refinement task. For subsequent task strategy, we employ two different reward models based on loss and CrystalBLEU metrics of code refinement. This reward-driven framework aims to produce more meaningful and useful review comments that optimize the rewards.

We conducted both quantitative and qualitative experiments to assess the relevance of the comments generated, focusing on BLEU scores and earned rewards for the different strategies implemented. Subsequently, we compare the performance of our best strategy against the state-of-the-art technique. Additionally, we investigate the usefulness of the generated comments compared to the baseline, utilizing \emph{GPT-4} as a judge. The evaluation results demonstrate the superiority of \oapp in generating more accurate and useful comments.

As part of future work, we aim to explore other reward models to further enhance the generation process. Further, we plan to adapt our framework to code refinement, to generate more accurate and meaningful code edits. We also intend to develop specific metrics to assess the generated comments, covering various aspects such as correctness, semantics, relevance, and usefulness for subsequent tasks.

\section*{Data availability}
We publicly release the replication package of our experiments online \cite{github_replication}.

\bibliographystyle{ACM-Reference-Format}
\bibliography{references}

%%% -*-BibTeX-*-
%%% Do NOT edit. File created by BibTeX with style
%%% ACM-Reference-Format-Journals [18-Jan-2012].

\begin{thebibliography}{90}

%%% ====================================================================
%%% NOTE TO THE USER: you can override these defaults by providing
%%% customized versions of any of these macros before the \bibliography
%%% command.  Each of them MUST provide its own final punctuation,
%%% except for \shownote{}, \showDOI{}, and \showURL{}.  The latter two
%%% do not use final punctuation, in order to avoid confusing it with
%%% the Web address.
%%%
%%% To suppress output of a particular field, define its macro to expand
%%% to an empty string, or better, \unskip, like this:
%%%
%%% \newcommand{\showDOI}[1]{\unskip}   % LaTeX syntax
%%%
%%% \def \showDOI #1{\unskip}           % plain TeX syntax
%%%
%%% ====================================================================

\ifx \showCODEN    \undefined \def \showCODEN     #1{\unskip}     \fi
\ifx \showDOI      \undefined \def \showDOI       #1{#1}\fi
\ifx \showISBNx    \undefined \def \showISBNx     #1{\unskip}     \fi
\ifx \showISBNxiii \undefined \def \showISBNxiii  #1{\unskip}     \fi
\ifx \showISSN     \undefined \def \showISSN      #1{\unskip}     \fi
\ifx \showLCCN     \undefined \def \showLCCN      #1{\unskip}     \fi
\ifx \shownote     \undefined \def \shownote      #1{#1}          \fi
\ifx \showarticletitle \undefined \def \showarticletitle #1{#1}   \fi
\ifx \showURL      \undefined \def \showURL       {\relax}        \fi
% The following commands are used for tagged output and should be
% invisible to TeX
\providecommand\bibfield[2]{#2}
\providecommand\bibinfo[2]{#2}
\providecommand\natexlab[1]{#1}
\providecommand\showeprint[2][]{arXiv:#2}

\bibitem[pmd(2000)]%
        {pmd}
 \bibinfo{year}{2000}\natexlab{}.
\newblock \bibinfo{title}{{PMD}}.
\newblock \bibinfo{howpublished}{\url{https://pmd.github.io/}}.
\newblock


\bibitem[fin(2005)]%
        {findBugs}
 \bibinfo{year}{2005}\natexlab{}.
\newblock \bibinfo{title}{{FindBugs}}.
\newblock \bibinfo{howpublished}{\url{https://findbugs.sourceforge.net/}}.
\newblock


\bibitem[Abo-eleneen et~al\mbox{.}(2023)]%
        {aboeleneen:ist23}
\bibfield{author}{\bibinfo{person}{Amr Abo-eleneen}, \bibinfo{person}{Ahammed Palliyali}, {and} \bibinfo{person}{Cagatay Catal}.} \bibinfo{year}{2023}\natexlab{}.
\newblock \showarticletitle{The role of Reinforcement Learning in software testing}.
\newblock \bibinfo{journal}{\emph{Information and Software Technology}}  \bibinfo{volume}{164} (\bibinfo{year}{2023}), \bibinfo{pages}{107325}.
\newblock
\urldef\tempurl%
\url{https://doi.org/10.1016/j.infsof.2023.107325}
\showDOI{\tempurl}


\bibitem[Ackerman et~al\mbox{.}(1989)]%
        {ackerman1989software}
\bibfield{author}{\bibinfo{person}{A.~Frank Ackerman}, \bibinfo{person}{Lynne~S. Buchwald}, {and} \bibinfo{person}{Frank~H. Lewski}.} \bibinfo{year}{1989}\natexlab{}.
\newblock \showarticletitle{Software inspections: an effective verification process}.
\newblock \bibinfo{journal}{\emph{IEEE software}} \bibinfo{volume}{6}, \bibinfo{number}{3} (\bibinfo{year}{1989}), \bibinfo{pages}{31--36}.
\newblock


\bibitem[Adamo et~al\mbox{.}(2018)]%
        {adamo:2018}
\bibfield{author}{\bibinfo{person}{David Adamo}, \bibinfo{person}{Md~Khorrom Khan}, \bibinfo{person}{Sreedevi Koppula}, {and} \bibinfo{person}{Ren{\'e}e Bryce}.} \bibinfo{year}{2018}\natexlab{}.
\newblock \showarticletitle{Reinforcement learning for android gui testing}. In \bibinfo{booktitle}{\emph{Proceedings of the 9th ACM SIGSOFT International Workshop on Automating TEST Case Design, Selection, and Evaluation}}. \bibinfo{pages}{2--8}.
\newblock


\bibitem[Ahmadi et~al\mbox{.}(2022)]%
        {ahmadi:2022}
\bibfield{author}{\bibinfo{person}{Hamidreza Ahmadi}, \bibinfo{person}{Mehrdad Ashtiani}, \bibinfo{person}{Mohammad~Abdollahi Azgomi}, {and} \bibinfo{person}{Raana Saheb-Nassagh}.} \bibinfo{year}{2022}\natexlab{}.
\newblock \showarticletitle{A DQN-based agent for automatic software refactoring}.
\newblock \bibinfo{journal}{\emph{Information and Software Technology}}  \bibinfo{volume}{147} (\bibinfo{year}{2022}), \bibinfo{pages}{106893}.
\newblock


\bibitem[Ahmed et~al\mbox{.}(2023)]%
        {ahmed:2023}
\bibfield{author}{\bibinfo{person}{Areeg Ahmed}, \bibinfo{person}{Shahira Azab}, {and} \bibinfo{person}{Yasser Abdelhamid}.} \bibinfo{year}{2023}\natexlab{}.
\newblock \showarticletitle{Source-Code Generation Using Deep Learning: A Survey}. In \bibinfo{booktitle}{\emph{Progress in Artificial Intelligence}}, \bibfield{editor}{\bibinfo{person}{Nuno Moniz}, \bibinfo{person}{Zita Vale}, \bibinfo{person}{Jos{\'e} Cascalho}, \bibinfo{person}{Catarina Silva}, {and} \bibinfo{person}{Raquel Sebasti{\~a}o}} (Eds.). \bibinfo{publisher}{Springer Nature Switzerland}, \bibinfo{address}{Cham}, \bibinfo{pages}{467--482}.
\newblock
\showISBNx{978-3-031-49011-8}


\bibitem[AI(2019)]%
        {berner:2019}
\bibfield{author}{\bibinfo{person}{Open AI}.} \bibinfo{year}{2019}\natexlab{}.
\newblock \showarticletitle{Dota 2 with large scale deep reinforcement learning}.
\newblock \bibinfo{journal}{\emph{arXiv preprint arXiv:1912.06680}} (\bibinfo{year}{2019}).
\newblock


\bibitem[Al-Zubaidi et~al\mbox{.}(2020)]%
        {Al-Zubaidi:2020}
\bibfield{author}{\bibinfo{person}{Wisam Haitham~Abbood Al-Zubaidi}, \bibinfo{person}{Patanamon Thongtanunam}, \bibinfo{person}{Hoa~Khanh Dam}, \bibinfo{person}{Chakkrit Tantithamthavorn}, {and} \bibinfo{person}{Aditya Ghose}.} \bibinfo{year}{2020}\natexlab{}.
\newblock \showarticletitle{Workload-Aware Reviewer Recommendation Using a Multi-Objective Search-Based Approach}. In \bibinfo{booktitle}{\emph{Proceedings of the 16th ACM International Conference on Predictive Models and Data Analytics in Software Engineering}} (Virtual, USA) \emph{(\bibinfo{series}{PROMISE 2020})}. \bibinfo{address}{New York, NY, USA}, \bibinfo{pages}{21–30}.
\newblock
\showISBNx{9781450381277}
\urldef\tempurl%
\url{https://doi.org/10.1145/3416508.3417115}
\showDOI{\tempurl}


\bibitem[{Ariyurek} et~al\mbox{.}(2019)]%
        {Ariyurek:2019}
\bibfield{author}{\bibinfo{person}{S. {Ariyurek}}, \bibinfo{person}{A. {Betin-Can}}, {and} \bibinfo{person}{E. {Surer}}.} \bibinfo{year}{2019}\natexlab{}.
\newblock \showarticletitle{Automated Video Game Testing Using Synthetic and Human-Like Agents}.
\newblock \bibinfo{journal}{\emph{IEEE Transactions on Games}} (\bibinfo{year}{2019}), \bibinfo{pages}{1--1}.
\newblock
\urldef\tempurl%
\url{https://doi.org/10.1109/TG.2019.2947597}
\showDOI{\tempurl}


\bibitem[Avgeriou et~al\mbox{.}(2016)]%
        {avgeriou2016managing}
\bibfield{author}{\bibinfo{person}{Paris Avgeriou}, \bibinfo{person}{Philippe Kruchten}, \bibinfo{person}{Ipek Ozkaya}, {and} \bibinfo{person}{Carolyn Seaman}.} \bibinfo{year}{2016}\natexlab{}.
\newblock \showarticletitle{Managing technical debt in software engineering ({D}agstuhl seminar 16162)}. In \bibinfo{booktitle}{\emph{Dagstuhl Reports}}, Vol.~\bibinfo{volume}{6}. Schloss Dagstuhl-Leibniz-Zentrum fuer Informatik.
\newblock


\bibitem[Bagherzadeh et~al\mbox{.}(2022)]%
        {bagherzadeh:tse22}
\bibfield{author}{\bibinfo{person}{Mojtaba Bagherzadeh}, \bibinfo{person}{Nafiseh Kahani}, {and} \bibinfo{person}{Lionel Briand}.} \bibinfo{year}{2022}\natexlab{}.
\newblock \showarticletitle{Reinforcement Learning for Test Case Prioritization}.
\newblock \bibinfo{journal}{\emph{IEEE Transactions on Software Engineering}} \bibinfo{volume}{48}, \bibinfo{number}{8} (\bibinfo{year}{2022}), \bibinfo{pages}{2836--2856}.
\newblock
\urldef\tempurl%
\url{https://doi.org/10.1109/TSE.2021.3070549}
\showDOI{\tempurl}


\bibitem[Bai et~al\mbox{.}(2022)]%
        {bai2022training}
\bibfield{author}{\bibinfo{person}{Yuntao Bai}, \bibinfo{person}{Andy Jones}, \bibinfo{person}{Kamal Ndousse}, \bibinfo{person}{Amanda Askell}, \bibinfo{person}{Anna Chen}, \bibinfo{person}{Nova DasSarma}, \bibinfo{person}{Dawn Drain}, \bibinfo{person}{Stanislav Fort}, \bibinfo{person}{Deep Ganguli}, \bibinfo{person}{Tom Henighan}, {et~al\mbox{.}}} \bibinfo{year}{2022}\natexlab{}.
\newblock \showarticletitle{Training a helpful and harmless assistant with reinforcement learning from human feedback}.
\newblock \bibinfo{journal}{\emph{arXiv preprint arXiv:2204.05862}} (\bibinfo{year}{2022}).
\newblock


\bibitem[Baker et~al\mbox{.}(2020)]%
        {Baker:2020}
\bibfield{author}{\bibinfo{person}{Bowen Baker}, \bibinfo{person}{Ingmar Kanitscheider}, \bibinfo{person}{Todor Markov}, \bibinfo{person}{Yi Wu}, \bibinfo{person}{Glenn Powell}, \bibinfo{person}{Bob McGrew}, {and} \bibinfo{person}{Igor Mordatch}.} \bibinfo{year}{2020}\natexlab{}.
\newblock \showarticletitle{Emergent Tool Use From Multi-Agent Autocurricula}.
\newblock \bibinfo{journal}{\emph{ArXiv}}  \bibinfo{volume}{abs/1909.07528} (\bibinfo{year}{2020}).
\newblock


\bibitem[Balachandran(2013)]%
        {balachandran2013reducing}
\bibfield{author}{\bibinfo{person}{Vipin Balachandran}.} \bibinfo{year}{2013}\natexlab{}.
\newblock \showarticletitle{Reducing human effort and improving quality in peer code reviews using automatic static analysis and reviewer recommendation}. In \bibinfo{booktitle}{\emph{2013 35th International Conference on Software Engineering (ICSE)}}. IEEE, \bibinfo{pages}{931--940}.
\newblock


\bibitem[Barriga et~al\mbox{.}(2020)]%
        {barriga:models2020}
\bibfield{author}{\bibinfo{person}{Angela Barriga}, \bibinfo{person}{Lawrence Mandow}, \bibinfo{person}{Jos\'{e} Luis~P\'{e}rez de~la Cruz}, \bibinfo{person}{Adrian Rutle}, \bibinfo{person}{Rogardt Heldal}, {and} \bibinfo{person}{Ludovico Iovino}.} \bibinfo{year}{2020}\natexlab{}.
\newblock \showarticletitle{A comparative study of reinforcement learning techniques to repair models}. In \bibinfo{booktitle}{\emph{Proceedings of the 23rd ACM/IEEE International Conference on Model Driven Engineering Languages and Systems: Companion Proceedings}} (Virtual Event, Canada) \emph{(\bibinfo{series}{MODELS '20})}. Article \bibinfo{articleno}{47}, \bibinfo{numpages}{9}~pages.
\newblock
\urldef\tempurl%
\url{https://doi.org/10.1145/3417990.3421395}
\showDOI{\tempurl}


\bibitem[Barto(2021)]%
        {barto2021reinforcement}
\bibfield{author}{\bibinfo{person}{Andrew~G Barto}.} \bibinfo{year}{2021}\natexlab{}.
\newblock \showarticletitle{Reinforcement learning: An introduction. by richard’s sutton}.
\newblock \bibinfo{journal}{\emph{SIAM Rev}} \bibinfo{volume}{6}, \bibinfo{number}{2} (\bibinfo{year}{2021}), \bibinfo{pages}{423}.
\newblock


\bibitem[Benjamini and Hochberg(1995)]%
        {benjamini1995controlling}
\bibfield{author}{\bibinfo{person}{Yoav Benjamini} {and} \bibinfo{person}{Yosef Hochberg}.} \bibinfo{year}{1995}\natexlab{}.
\newblock \showarticletitle{Controlling the false discovery rate: a practical and powerful approach to multiple testing}.
\newblock \bibinfo{journal}{\emph{Journal of the Royal statistical society: series B (Methodological)}} \bibinfo{volume}{57}, \bibinfo{number}{1} (\bibinfo{year}{1995}), \bibinfo{pages}{289--300}.
\newblock


\bibitem[{Bergdahl} et~al\mbox{.}(2020)]%
        {Bergdahl:cog20}
\bibfield{author}{\bibinfo{person}{J. {Bergdahl}}, \bibinfo{person}{C. {Gordillo}}, \bibinfo{person}{K. {Tollmar}}, {and} \bibinfo{person}{L. {Gisslén}}.} \bibinfo{year}{2020}\natexlab{}.
\newblock \showarticletitle{Augmenting Automated Game Testing with Deep Reinforcement Learning}. In \bibinfo{booktitle}{\emph{2020 IEEE Conference on Games (CoG)}}. \bibinfo{pages}{600--603}.
\newblock
\urldef\tempurl%
\url{https://doi.org/10.1109/CoG47356.2020.9231552}
\showDOI{\tempurl}


\bibitem[Bielik et~al\mbox{.}(2017)]%
        {bielik2017learning}
\bibfield{author}{\bibinfo{person}{Pavol Bielik}, \bibinfo{person}{Veselin Raychev}, {and} \bibinfo{person}{Martin Vechev}.} \bibinfo{year}{2017}\natexlab{}.
\newblock \showarticletitle{Learning a static analyzer from data}. In \bibinfo{booktitle}{\emph{International Conference on Computer Aided Verification}}. Springer, \bibinfo{pages}{233--253}.
\newblock


\bibitem[Bosu and Carver(2013)]%
        {bosu2013impact}
\bibfield{author}{\bibinfo{person}{Amiangshu Bosu} {and} \bibinfo{person}{Jeffrey~C Carver}.} \bibinfo{year}{2013}\natexlab{}.
\newblock \showarticletitle{Impact of peer code review on peer impression formation: A survey}. In \bibinfo{booktitle}{\emph{2013 ACM/IEEE International Symposium on Empirical Software Engineering and Measurement}}. IEEE, \bibinfo{pages}{133--142}.
\newblock


\bibitem[Carino and Andrews(2015)]%
        {carino:2015}
\bibfield{author}{\bibinfo{person}{Santo Carino} {and} \bibinfo{person}{James~H Andrews}.} \bibinfo{year}{2015}\natexlab{}.
\newblock \showarticletitle{Dynamically testing GUIs using ant colony optimization (t)}. In \bibinfo{booktitle}{\emph{2015 30th IEEE/ACM International Conference on Automated Software Engineering (ASE)}}. IEEE, \bibinfo{pages}{138--148}.
\newblock


\bibitem[Chen et~al\mbox{.}(2024)]%
        {chen:2024}
\bibfield{author}{\bibinfo{person}{Haoyang Chen}, \bibinfo{person}{Botong Xu}, {and} \bibinfo{person}{Kaiyang Zhong}.} \bibinfo{year}{2024}\natexlab{}.
\newblock \showarticletitle{Enhancing Software Effort Estimation through Reinforcement Learning-based Project Management-Oriented Feature Selection}.
\newblock \bibinfo{journal}{\emph{arXiv preprint arXiv:2403.16749}} (\bibinfo{year}{2024}).
\newblock


\bibitem[Chiang et~al\mbox{.}(2021)]%
        {chiang2021breaking}
\bibfield{author}{\bibinfo{person}{Ting-Rui Chiang}, \bibinfo{person}{Yi-Pei Chen}, \bibinfo{person}{Yi-Ting Yeh}, {and} \bibinfo{person}{Graham Neubig}.} \bibinfo{year}{2021}\natexlab{}.
\newblock \showarticletitle{Breaking down multilingual machine translation}.
\newblock \bibinfo{journal}{\emph{arXiv preprint arXiv:2110.08130}} (\bibinfo{year}{2021}).
\newblock


\bibitem[Cho and Behl(2020)]%
        {cho2020towards}
\bibfield{author}{\bibinfo{person}{Hyun~Jae Cho} {and} \bibinfo{person}{Madhur Behl}.} \bibinfo{year}{2020}\natexlab{}.
\newblock \showarticletitle{Towards automated safety coverage and testing for autonomous vehicles with reinforcement learning}.
\newblock \bibinfo{journal}{\emph{arXiv preprint arXiv:2005.13976}} (\bibinfo{year}{2020}).
\newblock


\bibitem[Chouchen et~al\mbox{.}(2021)]%
        {chouchen2021whoreview}
\bibfield{author}{\bibinfo{person}{Moataz Chouchen}, \bibinfo{person}{Ali Ouni}, \bibinfo{person}{Mohamed~Wiem Mkaouer}, \bibinfo{person}{Raula~Gaikovina Kula}, {and} \bibinfo{person}{Katsuro Inoue}.} \bibinfo{year}{2021}\natexlab{}.
\newblock \showarticletitle{WhoReview: A multi-objective search-based approach for code reviewers recommendation in modern code review}.
\newblock \bibinfo{journal}{\emph{Applied Soft Computing}}  \bibinfo{volume}{100} (\bibinfo{year}{2021}), \bibinfo{pages}{106908}.
\newblock


\bibitem[Collins et~al\mbox{.}(2021)]%
        {collins:deep2021}
\bibfield{author}{\bibinfo{person}{Eliane Collins}, \bibinfo{person}{Arilo Neto}, \bibinfo{person}{Auri Vincenzi}, {and} \bibinfo{person}{Jos{\'e} Maldonado}.} \bibinfo{year}{2021}\natexlab{}.
\newblock \showarticletitle{Deep reinforcement learning based android application gui testing}. In \bibinfo{booktitle}{\emph{Proceedings of the XXXV Brazilian Symposium on Software Engineering}}. \bibinfo{pages}{186--194}.
\newblock


\bibitem[Eghbali and Pradel(2022)]%
        {eghbali2022crystalbleu}
\bibfield{author}{\bibinfo{person}{Aryaz Eghbali} {and} \bibinfo{person}{Michael Pradel}.} \bibinfo{year}{2022}\natexlab{}.
\newblock \showarticletitle{CrystalBLEU: precisely and efficiently measuring the similarity of code}. In \bibinfo{booktitle}{\emph{Proceedings of the 37th IEEE/ACM International Conference on Automated Software Engineering}}. \bibinfo{pages}{1--12}.
\newblock


\bibitem[Eick et~al\mbox{.}(2001)]%
        {eick2001does}
\bibfield{author}{\bibinfo{person}{Stephen~G Eick}, \bibinfo{person}{Todd~L Graves}, \bibinfo{person}{Alan~F Karr}, \bibinfo{person}{J~Steve Marron}, {and} \bibinfo{person}{Audris Mockus}.} \bibinfo{year}{2001}\natexlab{}.
\newblock \showarticletitle{Does code decay? assessing the evidence from change management data}.
\newblock \bibinfo{journal}{\emph{IEEE Transactions on Software Engineering}} \bibinfo{volume}{27}, \bibinfo{number}{1} (\bibinfo{year}{2001}), \bibinfo{pages}{1--12}.
\newblock


\bibitem[Fagerland and Sandvik(2009)]%
        {fagerland2009wilcoxon}
\bibfield{author}{\bibinfo{person}{Morten~W Fagerland} {and} \bibinfo{person}{Leiv Sandvik}.} \bibinfo{year}{2009}\natexlab{}.
\newblock \showarticletitle{The wilcoxon--mann--whitney test under scrutiny}.
\newblock \bibinfo{journal}{\emph{Statistics in medicine}} \bibinfo{volume}{28}, \bibinfo{number}{10} (\bibinfo{year}{2009}), \bibinfo{pages}{1487--1497}.
\newblock


\bibitem[Feng et~al\mbox{.}(2023)]%
        {feng2023dense}
\bibfield{author}{\bibinfo{person}{Shuo Feng}, \bibinfo{person}{Haowei Sun}, \bibinfo{person}{Xintao Yan}, \bibinfo{person}{Haojie Zhu}, \bibinfo{person}{Zhengxia Zou}, \bibinfo{person}{Shengyin Shen}, {and} \bibinfo{person}{Henry~X Liu}.} \bibinfo{year}{2023}\natexlab{}.
\newblock \showarticletitle{Dense reinforcement learning for safety validation of autonomous vehicles}.
\newblock \bibinfo{journal}{\emph{Nature}} \bibinfo{volume}{615}, \bibinfo{number}{7953} (\bibinfo{year}{2023}), \bibinfo{pages}{620--627}.
\newblock


\bibitem[Fowler and Foemmel(2006)]%
        {fowler2006continuous}
\bibfield{author}{\bibinfo{person}{Martin Fowler} {and} \bibinfo{person}{Matthew Foemmel}.} \bibinfo{year}{2006}\natexlab{}.
\newblock \bibinfo{title}{Continuous integration}.
\newblock \bibinfo{howpublished}{\url{https://martinfowler.com/articles/continuousIntegration.html}}.
\newblock


\bibitem[Gao et~al\mbox{.}(2023)]%
        {gao2023scaling}
\bibfield{author}{\bibinfo{person}{Leo Gao}, \bibinfo{person}{John Schulman}, {and} \bibinfo{person}{Jacob Hilton}.} \bibinfo{year}{2023}\natexlab{}.
\newblock \showarticletitle{Scaling laws for reward model overoptimization}. In \bibinfo{booktitle}{\emph{International Conference on Machine Learning}}. PMLR, \bibinfo{pages}{10835--10866}.
\newblock


\bibitem[Glaese et~al\mbox{.}(2022)]%
        {glaese2022improving}
\bibfield{author}{\bibinfo{person}{Amelia Glaese}, \bibinfo{person}{Nat McAleese}, \bibinfo{person}{Maja Tr{\k{e}}bacz}, \bibinfo{person}{John Aslanides}, \bibinfo{person}{Vlad Firoiu}, \bibinfo{person}{Timo Ewalds}, \bibinfo{person}{Maribeth Rauh}, \bibinfo{person}{Laura Weidinger}, \bibinfo{person}{Martin Chadwick}, \bibinfo{person}{Phoebe Thacker}, {et~al\mbox{.}}} \bibinfo{year}{2022}\natexlab{}.
\newblock \showarticletitle{Improving alignment of dialogue agents via targeted human judgements}.
\newblock \bibinfo{journal}{\emph{arXiv preprint arXiv:2209.14375}} (\bibinfo{year}{2022}).
\newblock


\bibitem[Goodfellow et~al\mbox{.}(2020)]%
        {goodfellow2020generative}
\bibfield{author}{\bibinfo{person}{Ian Goodfellow}, \bibinfo{person}{Jean Pouget-Abadie}, \bibinfo{person}{Mehdi Mirza}, \bibinfo{person}{Bing Xu}, \bibinfo{person}{David Warde-Farley}, \bibinfo{person}{Sherjil Ozair}, \bibinfo{person}{Aaron Courville}, {and} \bibinfo{person}{Yoshua Bengio}.} \bibinfo{year}{2020}\natexlab{}.
\newblock \showarticletitle{Generative adversarial networks}.
\newblock \bibinfo{journal}{\emph{Commun. ACM}} \bibinfo{volume}{63}, \bibinfo{number}{11} (\bibinfo{year}{2020}), \bibinfo{pages}{139--144}.
\newblock


\bibitem[Gupta and Sundaresan(2018)]%
        {gupta2018intelligent}
\bibfield{author}{\bibinfo{person}{Anshul Gupta} {and} \bibinfo{person}{Neel Sundaresan}.} \bibinfo{year}{2018}\natexlab{}.
\newblock \showarticletitle{Intelligent code reviews using deep learning}. In \bibinfo{booktitle}{\emph{Proceedings of the 24th ACM SIGKDD International Conference on Knowledge Discovery and Data Mining (KDD’18) Deep Learning Day}}.
\newblock


\bibitem[Haouari et~al\mbox{.}(2023)]%
        {haouari:2023}
\bibfield{author}{\bibinfo{person}{Bakhta Haouari}, \bibinfo{person}{Rania Mzid}, {and} \bibinfo{person}{Olfa Mosbahi}.} \bibinfo{year}{2023}\natexlab{}.
\newblock \showarticletitle{On the Use of Reinforcement Learning for Real-Time System Design and Refactoring}. In \bibinfo{booktitle}{\emph{Intelligent Systems Design and Applications}}, \bibfield{editor}{\bibinfo{person}{Ajith Abraham}, \bibinfo{person}{Sabri Pllana}, \bibinfo{person}{Gabriella Casalino}, \bibinfo{person}{Kun Ma}, {and} \bibinfo{person}{Anu Bajaj}} (Eds.). \bibinfo{publisher}{Springer Nature Switzerland}, \bibinfo{address}{Cham}, \bibinfo{pages}{503--512}.
\newblock
\showISBNx{978-3-031-35501-1}


\bibitem[Harries et~al\mbox{.}(2020)]%
        {harries:drift2020}
\bibfield{author}{\bibinfo{person}{Luke Harries}, \bibinfo{person}{Rebekah~Storan Clarke}, \bibinfo{person}{Timothy Chapman}, \bibinfo{person}{Swamy~VPLN Nallamalli}, \bibinfo{person}{Levent Ozgur}, \bibinfo{person}{Shuktika Jain}, \bibinfo{person}{Alex Leung}, \bibinfo{person}{Steve Lim}, \bibinfo{person}{Aaron Dietrich}, \bibinfo{person}{Jos{\'e}~Miguel Hern{\'a}ndez-Lobato}, {et~al\mbox{.}}} \bibinfo{year}{2020}\natexlab{}.
\newblock \showarticletitle{Drift: Deep reinforcement learning for functional software testing}.
\newblock \bibinfo{journal}{\emph{arXiv preprint arXiv:2007.08220}} (\bibinfo{year}{2020}).
\newblock


\bibitem[Hessel et~al\mbox{.}(2018)]%
        {Hessel2018}
\bibfield{author}{\bibinfo{person}{Matteo Hessel}, \bibinfo{person}{Joseph Modayil}, \bibinfo{person}{H.~V. Hasselt}, \bibinfo{person}{Tom Schaul}, \bibinfo{person}{Georg Ostrovski}, \bibinfo{person}{Will Dabney}, \bibinfo{person}{Dan Horgan}, \bibinfo{person}{Bilal Piot}, \bibinfo{person}{Mohammad~Gheshlaghi Azar}, {and} \bibinfo{person}{David Silver}.} \bibinfo{year}{2018}\natexlab{}.
\newblock \showarticletitle{Rainbow: Combining Improvements in Deep Reinforcement Learning}. In \bibinfo{booktitle}{\emph{AAAI}}.
\newblock


\bibitem[Hinton et~al\mbox{.}(2015)]%
        {hinton2015distilling}
\bibfield{author}{\bibinfo{person}{Geoffrey Hinton}, \bibinfo{person}{Oriol Vinyals}, {and} \bibinfo{person}{Jeff Dean}.} \bibinfo{year}{2015}\natexlab{}.
\newblock \showarticletitle{Distilling the knowledge in a neural network}.
\newblock \bibinfo{journal}{\emph{arXiv preprint arXiv:1503.02531}} (\bibinfo{year}{2015}).
\newblock


\bibitem[Hong et~al\mbox{.}(2022)]%
        {hong2022commentfinder}
\bibfield{author}{\bibinfo{person}{Yang Hong}, \bibinfo{person}{Chakkrit Tantithamthavorn}, \bibinfo{person}{Patanamon Thongtanunam}, {and} \bibinfo{person}{Aldeida Aleti}.} \bibinfo{year}{2022}\natexlab{}.
\newblock \showarticletitle{Commentfinder: a simpler, faster, more accurate code review comments recommendation}. In \bibinfo{booktitle}{\emph{Proceedings of the 30th ACM joint European software engineering conference and symposium on the foundations of software engineering}}. \bibinfo{pages}{507--519}.
\newblock


\bibitem[Hu et~al\mbox{.}(2021)]%
        {hu2021lora}
\bibfield{author}{\bibinfo{person}{Edward~J Hu}, \bibinfo{person}{Yelong Shen}, \bibinfo{person}{Phillip Wallis}, \bibinfo{person}{Zeyuan Allen-Zhu}, \bibinfo{person}{Yuanzhi Li}, \bibinfo{person}{Shean Wang}, \bibinfo{person}{Lu Wang}, {and} \bibinfo{person}{Weizhu Chen}.} \bibinfo{year}{2021}\natexlab{}.
\newblock \showarticletitle{Lora: Low-rank adaptation of large language models}.
\newblock \bibinfo{journal}{\emph{arXiv preprint arXiv:2106.09685}} (\bibinfo{year}{2021}).
\newblock


\bibitem[Joyce(2011)]%
        {joyce2011kullback}
\bibfield{author}{\bibinfo{person}{James~M Joyce}.} \bibinfo{year}{2011}\natexlab{}.
\newblock \showarticletitle{Kullback-leibler divergence}.
\newblock In \bibinfo{booktitle}{\emph{International encyclopedia of statistical science}}. \bibinfo{publisher}{Springer}, \bibinfo{pages}{720--722}.
\newblock


\bibitem[Kim et~al\mbox{.}(2021)]%
        {kim2021comparing}
\bibfield{author}{\bibinfo{person}{Taehyeon Kim}, \bibinfo{person}{Jaehoon Oh}, \bibinfo{person}{NakYil Kim}, \bibinfo{person}{Sangwook Cho}, {and} \bibinfo{person}{Se-Young Yun}.} \bibinfo{year}{2021}\natexlab{}.
\newblock \showarticletitle{Comparing kullback-leibler divergence and mean squared error loss in knowledge distillation}.
\newblock \bibinfo{journal}{\emph{arXiv preprint arXiv:2105.08919}} (\bibinfo{year}{2021}).
\newblock


\bibitem[Koren et~al\mbox{.}(2018)]%
        {koren2018adaptive}
\bibfield{author}{\bibinfo{person}{Mark Koren}, \bibinfo{person}{Saud Alsaif}, \bibinfo{person}{Ritchie Lee}, {and} \bibinfo{person}{Mykel~J Kochenderfer}.} \bibinfo{year}{2018}\natexlab{}.
\newblock \showarticletitle{Adaptive stress testing for autonomous vehicles}. In \bibinfo{booktitle}{\emph{2018 IEEE Intelligent Vehicles Symposium (IV)}}. IEEE, \bibinfo{pages}{1--7}.
\newblock


\bibitem[Landis and Koch(1977)]%
        {landis1977measurement}
\bibfield{author}{\bibinfo{person}{J~Richard Landis} {and} \bibinfo{person}{Gary~G Koch}.} \bibinfo{year}{1977}\natexlab{}.
\newblock \showarticletitle{The measurement of observer agreement for categorical data}.
\newblock \bibinfo{journal}{\emph{biometrics}} (\bibinfo{year}{1977}), \bibinfo{pages}{159--174}.
\newblock


\bibitem[Le et~al\mbox{.}(2022)]%
        {le2022coderl}
\bibfield{author}{\bibinfo{person}{Hung Le}, \bibinfo{person}{Yue Wang}, \bibinfo{person}{Akhilesh~Deepak Gotmare}, \bibinfo{person}{Silvio Savarese}, {and} \bibinfo{person}{Steven Chu~Hong Hoi}.} \bibinfo{year}{2022}\natexlab{}.
\newblock \showarticletitle{Coderl: Mastering code generation through pretrained models and deep reinforcement learning}.
\newblock \bibinfo{journal}{\emph{Advances in Neural Information Processing Systems}}  \bibinfo{volume}{35} (\bibinfo{year}{2022}), \bibinfo{pages}{21314--21328}.
\newblock


\bibitem[Lee et~al\mbox{.}(2023)]%
        {lee2023rlaif}
\bibfield{author}{\bibinfo{person}{Harrison Lee}, \bibinfo{person}{Samrat Phatale}, \bibinfo{person}{Hassan Mansoor}, \bibinfo{person}{Kellie~Ren Lu}, \bibinfo{person}{Thomas Mesnard}, \bibinfo{person}{Johan Ferret}, \bibinfo{person}{Colton Bishop}, \bibinfo{person}{Ethan Hall}, \bibinfo{person}{Victor Carbune}, {and} \bibinfo{person}{Abhinav Rastogi}.} \bibinfo{year}{2023}\natexlab{}.
\newblock \showarticletitle{Rlaif: Scaling reinforcement learning from human feedback with ai feedback}.
\newblock \bibinfo{journal}{\emph{arXiv e-prints}} (\bibinfo{year}{2023}).
\newblock


\bibitem[Lee et~al\mbox{.}(2017)]%
        {Lee:fse2017}
\bibfield{author}{\bibinfo{person}{Sun-Ro Lee}, \bibinfo{person}{Min-Jae Heo}, \bibinfo{person}{Chan-Gun Lee}, \bibinfo{person}{Milhan Kim}, {and} \bibinfo{person}{Gaeul Jeong}.} \bibinfo{year}{2017}\natexlab{}.
\newblock \showarticletitle{Applying Deep Learning Based Automatic Bug Triager to Industrial Projects}. In \bibinfo{booktitle}{\emph{Proceedings of the 2017 11th Joint Meeting on Foundations of Software Engineering}} (Paderborn, Germany) \emph{(\bibinfo{series}{ESEC/FSE 2017})}. \bibinfo{pages}{926–931}.
\newblock
\showISBNx{9781450351058}
\urldef\tempurl%
\url{https://doi.org/10.1145/3106237.3117776}
\showDOI{\tempurl}


\bibitem[Li et~al\mbox{.}(2023a)]%
        {li2023code}
\bibfield{author}{\bibinfo{person}{Ruiyin Li}, \bibinfo{person}{Peng Liang}, {and} \bibinfo{person}{Paris Avgeriou}.} \bibinfo{year}{2023}\natexlab{a}.
\newblock \showarticletitle{Code reviewer recommendation for architecture violations: An exploratory study}. In \bibinfo{booktitle}{\emph{Proceedings of the 27th International Conference on Evaluation and Assessment in Software Engineering}}. \bibinfo{pages}{42--51}.
\newblock


\bibitem[Li et~al\mbox{.}(2023b)]%
        {li2023alpacaeval}
\bibfield{author}{\bibinfo{person}{Xuechen Li}, \bibinfo{person}{Tianyi Zhang}, \bibinfo{person}{Yann Dubois}, \bibinfo{person}{Rohan Taori}, \bibinfo{person}{Ishaan Gulrajani}, \bibinfo{person}{Carlos Guestrin}, \bibinfo{person}{Percy Liang}, {and} \bibinfo{person}{Tatsunori~B Hashimoto}.} \bibinfo{year}{2023}\natexlab{b}.
\newblock \bibinfo{title}{Alpacaeval: An automatic evaluator of instruction-following models}.
\newblock
\newblock


\bibitem[Li et~al\mbox{.}(2022)]%
        {li2022automating}
\bibfield{author}{\bibinfo{person}{Zhiyu Li}, \bibinfo{person}{Shuai Lu}, \bibinfo{person}{Daya Guo}, \bibinfo{person}{Nan Duan}, \bibinfo{person}{Shailesh Jannu}, \bibinfo{person}{Grant Jenks}, \bibinfo{person}{Deep Majumder}, \bibinfo{person}{Jared Green}, \bibinfo{person}{Alexey Svyatkovskiy}, \bibinfo{person}{Shengyu Fu}, {et~al\mbox{.}}} \bibinfo{year}{2022}\natexlab{}.
\newblock \showarticletitle{Automating code review activities by large-scale pre-training}. In \bibinfo{booktitle}{\emph{Proceedings of the 30th ACM Joint European Software Engineering Conference and Symposium on the Foundations of Software Engineering}}. \bibinfo{pages}{1035--1047}.
\newblock


\bibitem[Li et~al\mbox{.}(2017)]%
        {li2017automatic}
\bibfield{author}{\bibinfo{person}{Zhixing Li}, \bibinfo{person}{Yue Yu}, \bibinfo{person}{Gang Yin}, \bibinfo{person}{Tao Wang}, \bibinfo{person}{Qiang Fan}, {and} \bibinfo{person}{Huaimin Wang}.} \bibinfo{year}{2017}\natexlab{}.
\newblock \showarticletitle{Automatic Classification of Review Comments in Pull-based Development Model.}. In \bibinfo{booktitle}{\emph{SEKE}}. \bibinfo{pages}{572--577}.
\newblock


\bibitem[Macbeth et~al\mbox{.}(2011)]%
        {macbeth2011cliff}
\bibfield{author}{\bibinfo{person}{Guillermo Macbeth}, \bibinfo{person}{Eugenia Razumiejczyk}, {and} \bibinfo{person}{Rub{\'e}n~Daniel Ledesma}.} \bibinfo{year}{2011}\natexlab{}.
\newblock \showarticletitle{Cliff's Delta Calculator: A non-parametric effect size program for two groups of observations}.
\newblock \bibinfo{journal}{\emph{Universitas Psychologica}} \bibinfo{volume}{10}, \bibinfo{number}{2} (\bibinfo{year}{2011}), \bibinfo{pages}{545--555}.
\newblock


\bibitem[Mariani et~al\mbox{.}(2012)]%
        {mariani:2012}
\bibfield{author}{\bibinfo{person}{Leonardo Mariani}, \bibinfo{person}{Mauro Pezze}, \bibinfo{person}{Oliviero Riganelli}, {and} \bibinfo{person}{Mauro Santoro}.} \bibinfo{year}{2012}\natexlab{}.
\newblock \showarticletitle{Autoblacktest: Automatic black-box testing of interactive applications}. In \bibinfo{booktitle}{\emph{2012 IEEE fifth international conference on software testing, verification and validation}}. IEEE, \bibinfo{pages}{81--90}.
\newblock


\bibitem[McIntosh et~al\mbox{.}(2014)]%
        {mcintosh2014impact}
\bibfield{author}{\bibinfo{person}{Shane McIntosh}, \bibinfo{person}{Yasutaka Kamei}, \bibinfo{person}{Bram Adams}, {and} \bibinfo{person}{Ahmed~E Hassan}.} \bibinfo{year}{2014}\natexlab{}.
\newblock \showarticletitle{The impact of code review coverage and code review participation on software quality: A case study of the {Qt}, {VTK}, and {ITK} projects}. In \bibinfo{booktitle}{\emph{11th working conference on mining software repositories}}. \bibinfo{pages}{192--201}.
\newblock


\bibitem[McIntosh et~al\mbox{.}(2016)]%
        {mcintosh2016empirical}
\bibfield{author}{\bibinfo{person}{Shane McIntosh}, \bibinfo{person}{Yasutaka Kamei}, \bibinfo{person}{Bram Adams}, {and} \bibinfo{person}{Ahmed~E Hassan}.} \bibinfo{year}{2016}\natexlab{}.
\newblock \showarticletitle{An empirical study of the impact of modern code review practices on software quality}.
\newblock \bibinfo{journal}{\emph{Empirical Software Engineering}} \bibinfo{volume}{21}, \bibinfo{number}{5} (\bibinfo{year}{2016}), \bibinfo{pages}{2146--2189}.
\newblock


\bibitem[Mnih et~al\mbox{.}(2013)]%
        {Mnih2013}
\bibfield{author}{\bibinfo{person}{Volodymyr Mnih}, \bibinfo{person}{Koray Kavukcuoglu}, \bibinfo{person}{David Silver}, \bibinfo{person}{Alex Graves}, \bibinfo{person}{Ioannis Antonoglou}, \bibinfo{person}{Daan Wierstra}, {and} \bibinfo{person}{Martin~A. Riedmiller}.} \bibinfo{year}{2013}\natexlab{}.
\newblock \showarticletitle{Playing Atari with Deep Reinforcement Learning}.
\newblock \bibinfo{journal}{\emph{ArXiv}}  \bibinfo{volume}{abs/1312.5602} (\bibinfo{year}{2013}).
\newblock


\bibitem[Mnih et~al\mbox{.}(2015)]%
        {Mnih2015}
\bibfield{author}{\bibinfo{person}{Volodymyr Mnih}, \bibinfo{person}{Koray Kavukcuoglu}, \bibinfo{person}{David Silver}, \bibinfo{person}{Andrei~A. Rusu}, \bibinfo{person}{Joel Veness}, \bibinfo{person}{Marc~G. Bellemare}, \bibinfo{person}{Alex Graves}, \bibinfo{person}{Martin~A. Riedmiller}, \bibinfo{person}{Andreas Fidjeland}, \bibinfo{person}{Georg Ostrovski}, \bibinfo{person}{Stig Petersen}, \bibinfo{person}{Charlie Beattie}, \bibinfo{person}{Amir Sadik}, \bibinfo{person}{Ioannis Antonoglou}, \bibinfo{person}{Helen King}, \bibinfo{person}{Dharshan Kumaran}, \bibinfo{person}{Daan Wierstra}, \bibinfo{person}{Shane Legg}, {and} \bibinfo{person}{Demis Hassabis}.} \bibinfo{year}{2015}\natexlab{}.
\newblock \showarticletitle{Human-level control through deep reinforcement learning}.
\newblock \bibinfo{journal}{\emph{Nature}}  \bibinfo{volume}{518} (\bibinfo{year}{2015}), \bibinfo{pages}{529--533}.
\newblock


\bibitem[Morales et~al\mbox{.}(2015)]%
        {morales2015code}
\bibfield{author}{\bibinfo{person}{Rodrigo Morales}, \bibinfo{person}{Shane McIntosh}, {and} \bibinfo{person}{Foutse Khomh}.} \bibinfo{year}{2015}\natexlab{}.
\newblock \showarticletitle{Do code review practices impact design quality? A case study of the {Qt}, {VTK}, and {ITK} projects}. In \bibinfo{booktitle}{\emph{2015 IEEE 22nd international conference on software analysis, evolution, and reengineering (SANER)}}. IEEE, \bibinfo{pages}{171--180}.
\newblock


\bibitem[Ouyang et~al\mbox{.}(2022)]%
        {ouyang2022training}
\bibfield{author}{\bibinfo{person}{Long Ouyang}, \bibinfo{person}{Jeffrey Wu}, \bibinfo{person}{Xu Jiang}, \bibinfo{person}{Diogo Almeida}, \bibinfo{person}{Carroll Wainwright}, \bibinfo{person}{Pamela Mishkin}, \bibinfo{person}{Chong Zhang}, \bibinfo{person}{Sandhini Agarwal}, \bibinfo{person}{Katarina Slama}, \bibinfo{person}{Alex Ray}, {et~al\mbox{.}}} \bibinfo{year}{2022}\natexlab{}.
\newblock \showarticletitle{Training language models to follow instructions with human feedback}.
\newblock \bibinfo{journal}{\emph{Advances in neural information processing systems}}  \bibinfo{volume}{35} (\bibinfo{year}{2022}), \bibinfo{pages}{27730--27744}.
\newblock


\bibitem[Papineni et~al\mbox{.}(2002)]%
        {papineni2002bleu}
\bibfield{author}{\bibinfo{person}{Kishore Papineni}, \bibinfo{person}{Salim Roukos}, \bibinfo{person}{Todd Ward}, {and} \bibinfo{person}{Wei-Jing Zhu}.} \bibinfo{year}{2002}\natexlab{}.
\newblock \showarticletitle{Bleu: a method for automatic evaluation of machine translation}. In \bibinfo{booktitle}{\emph{Proceedings of the 40th annual meeting of the Association for Computational Linguistics}}. \bibinfo{pages}{311--318}.
\newblock


\bibitem[Reimers and Gurevych(2019)]%
        {reimers2019sentence}
\bibfield{author}{\bibinfo{person}{Nils Reimers} {and} \bibinfo{person}{Iryna Gurevych}.} \bibinfo{year}{2019}\natexlab{}.
\newblock \showarticletitle{Sentence-bert: Sentence embeddings using siamese bert-networks}.
\newblock \bibinfo{journal}{\emph{arXiv preprint arXiv:1908.10084}} (\bibinfo{year}{2019}).
\newblock


\bibitem[Replication Package({[n.\,d.]})]%
        {github_replication}
Replication Package \bibinfo{year}{[n.\,d.]}\natexlab{}.
\newblock \bibinfo{title}{Replication Package}.
\newblock \bibinfo{howpublished}{\url{https://github.com/OussamaSghaier/RL4CR}}.
\newblock


\bibitem[Roziere et~al\mbox{.}(2023)]%
        {roziere2023code}
\bibfield{author}{\bibinfo{person}{Baptiste Roziere}, \bibinfo{person}{Jonas Gehring}, \bibinfo{person}{Fabian Gloeckle}, \bibinfo{person}{Sten Sootla}, \bibinfo{person}{Itai Gat}, \bibinfo{person}{Xiaoqing~Ellen Tan}, \bibinfo{person}{Yossi Adi}, \bibinfo{person}{Jingyu Liu}, \bibinfo{person}{Tal Remez}, \bibinfo{person}{J{\'e}r{\'e}my Rapin}, {et~al\mbox{.}}} \bibinfo{year}{2023}\natexlab{}.
\newblock \showarticletitle{Code llama: Open foundation models for code}.
\newblock \bibinfo{journal}{\emph{arXiv preprint arXiv:2308.12950}} (\bibinfo{year}{2023}).
\newblock


\bibitem[Sadowski et~al\mbox{.}(2015)]%
        {sadowski2015tricorder}
\bibfield{author}{\bibinfo{person}{Caitlin Sadowski}, \bibinfo{person}{Jeffrey Van~Gogh}, \bibinfo{person}{Ciera Jaspan}, \bibinfo{person}{Emma Soderberg}, {and} \bibinfo{person}{Collin Winter}.} \bibinfo{year}{2015}\natexlab{}.
\newblock \showarticletitle{Tricorder: Building a program analysis ecosystem}. In \bibinfo{booktitle}{\emph{2015 IEEE/ACM 37th IEEE International Conference on Software Engineering}}, Vol.~\bibinfo{volume}{1}. IEEE, \bibinfo{pages}{598--608}.
\newblock


\bibitem[Schulman et~al\mbox{.}(2017)]%
        {schulman2017proximal}
\bibfield{author}{\bibinfo{person}{John Schulman}, \bibinfo{person}{Filip Wolski}, \bibinfo{person}{Prafulla Dhariwal}, \bibinfo{person}{Alec Radford}, {and} \bibinfo{person}{Oleg Klimov}.} \bibinfo{year}{2017}\natexlab{}.
\newblock \showarticletitle{Proximal policy optimization algorithms}.
\newblock \bibinfo{journal}{\emph{arXiv preprint arXiv:1707.06347}} (\bibinfo{year}{2017}).
\newblock


\bibitem[Sennrich et~al\mbox{.}(2015)]%
        {sennrich2015neural}
\bibfield{author}{\bibinfo{person}{Rico Sennrich}, \bibinfo{person}{Barry Haddow}, {and} \bibinfo{person}{Alexandra Birch}.} \bibinfo{year}{2015}\natexlab{}.
\newblock \showarticletitle{Neural machine translation of rare words with subword units}.
\newblock \bibinfo{journal}{\emph{arXiv preprint arXiv:1508.07909}} (\bibinfo{year}{2015}).
\newblock


\bibitem[Sghaier and Sahraoui(2023)]%
        {sghaier2023multi}
\bibfield{author}{\bibinfo{person}{Oussama~Ben Sghaier} {and} \bibinfo{person}{Houari Sahraoui}.} \bibinfo{year}{2023}\natexlab{}.
\newblock \showarticletitle{A Multi-Step Learning Approach to Assist Code Review}. In \bibinfo{booktitle}{\emph{2023 IEEE 23rd International Conference on Software Analysis, Evolution, and Reengineering (SANER)}}. IEEE.
\newblock


\bibitem[Sghaier and Sahraoui(2024)]%
        {sghaier2024improving}
\bibfield{author}{\bibinfo{person}{Oussama~Ben Sghaier} {and} \bibinfo{person}{Houari Sahraoui}.} \bibinfo{year}{2024}\natexlab{}.
\newblock \showarticletitle{Improving the Learning of Code Review Successive Tasks with Cross-Task Knowledge Distillation}.
\newblock \bibinfo{journal}{\emph{arXiv preprint arXiv:2402.02063}} (\bibinfo{year}{2024}).
\newblock


\bibitem[Shahin et~al\mbox{.}(2017)]%
        {shahin2017continuous}
\bibfield{author}{\bibinfo{person}{Mojtaba Shahin}, \bibinfo{person}{Muhammad~Ali Babar}, {and} \bibinfo{person}{Liming Zhu}.} \bibinfo{year}{2017}\natexlab{}.
\newblock \showarticletitle{Continuous integration, delivery and deployment: a systematic review on approaches, tools, challenges and practices}.
\newblock \bibinfo{journal}{\emph{IEEE Access}}  \bibinfo{volume}{5} (\bibinfo{year}{2017}), \bibinfo{pages}{3909--3943}.
\newblock


\bibitem[Siow et~al\mbox{.}(2020)]%
        {siow2020core}
\bibfield{author}{\bibinfo{person}{Jing~Kai Siow}, \bibinfo{person}{Cuiyun Gao}, \bibinfo{person}{Lingling Fan}, \bibinfo{person}{Sen Chen}, {and} \bibinfo{person}{Yang Liu}.} \bibinfo{year}{2020}\natexlab{}.
\newblock \showarticletitle{Core: Automating review recommendation for code changes}. In \bibinfo{booktitle}{\emph{2020 IEEE 27th International Conference on Software Analysis, Evolution and Reengineering (SANER)}}. IEEE, \bibinfo{pages}{284--295}.
\newblock


\bibitem[Sukur et~al\mbox{.}(2024)]%
        {sukur:sc2024}
\bibfield{author}{\bibinfo{person}{Nata{\v{s}}a Sukur}, \bibinfo{person}{Nemanja Milo{\v{s}}evi{\'c}}, \bibinfo{person}{Doni Pracner}, {and} \bibinfo{person}{Zoran Budimac}.} \bibinfo{year}{2024}\natexlab{}.
\newblock \showarticletitle{Automated program improvement with reinforcement learning and graph neural networks}.
\newblock \bibinfo{journal}{\emph{Soft Computing}} \bibinfo{volume}{28}, \bibinfo{number}{3} (\bibinfo{year}{2024}), \bibinfo{pages}{2593--2604}.
\newblock


\bibitem[Thongtanunam et~al\mbox{.}(2015)]%
        {thongtanunam2015should}
\bibfield{author}{\bibinfo{person}{Patanamon Thongtanunam}, \bibinfo{person}{Chakkrit Tantithamthavorn}, \bibinfo{person}{Raula~Gaikovina Kula}, \bibinfo{person}{Norihiro Yoshida}, \bibinfo{person}{Hajimu Iida}, {and} \bibinfo{person}{Ken-ichi Matsumoto}.} \bibinfo{year}{2015}\natexlab{}.
\newblock \showarticletitle{Who should review my code? a file location-based code-reviewer recommendation approach for modern code review}. In \bibinfo{booktitle}{\emph{2015 IEEE 22nd International Conference on Software Analysis, Evolution, and Reengineering (SANER)}}. IEEE, \bibinfo{pages}{141--150}.
\newblock


\bibitem[Tlili and Chikhi(2021)]%
        {tlili:2021}
\bibfield{author}{\bibinfo{person}{Ahmed Tlili} {and} \bibinfo{person}{Salim Chikhi}.} \bibinfo{year}{2021}\natexlab{}.
\newblock \showarticletitle{Risks analyzing and management in software project management using fuzzy cognitive maps with reinforcement learning}.
\newblock \bibinfo{journal}{\emph{Informatica}} \bibinfo{volume}{45}, \bibinfo{number}{1} (\bibinfo{year}{2021}).
\newblock


\bibitem[Touvron et~al\mbox{.}(2023)]%
        {touvron2023llama}
\bibfield{author}{\bibinfo{person}{Hugo Touvron}, \bibinfo{person}{Louis Martin}, \bibinfo{person}{Kevin Stone}, \bibinfo{person}{Peter Albert}, \bibinfo{person}{Amjad Almahairi}, \bibinfo{person}{Yasmine Babaei}, \bibinfo{person}{Nikolay Bashlykov}, \bibinfo{person}{Soumya Batra}, \bibinfo{person}{Prajjwal Bhargava}, \bibinfo{person}{Shruti Bhosale}, {et~al\mbox{.}}} \bibinfo{year}{2023}\natexlab{}.
\newblock \showarticletitle{Llama 2: Open foundation and fine-tuned chat models}.
\newblock \bibinfo{journal}{\emph{arXiv preprint arXiv:2307.09288}} (\bibinfo{year}{2023}).
\newblock


\bibitem[Tufan et~al\mbox{.}(2021)]%
        {tufan2021towards}
\bibfield{author}{\bibinfo{person}{Rosalia Tufan}, \bibinfo{person}{Luca Pascarella}, \bibinfo{person}{Michele Tufanoy}, \bibinfo{person}{Denys Poshyvanykz}, {and} \bibinfo{person}{Gabriele Bavota}.} \bibinfo{year}{2021}\natexlab{}.
\newblock \showarticletitle{Towards automating code review activities}. In \bibinfo{booktitle}{\emph{2021 IEEE/ACM 43rd International Conference on Software Engineering (ICSE)}}. \bibinfo{pages}{163--174}.
\newblock


\bibitem[Tufano et~al\mbox{.}(2022a)]%
        {tufano2022using}
\bibfield{author}{\bibinfo{person}{Rosalia Tufano}, \bibinfo{person}{Simone Masiero}, \bibinfo{person}{Antonio Mastropaolo}, \bibinfo{person}{Luca Pascarella}, \bibinfo{person}{Denys Poshyvanyk}, {and} \bibinfo{person}{Gabriele Bavota}.} \bibinfo{year}{2022}\natexlab{a}.
\newblock \showarticletitle{Using pre-trained models to boost code review automation}.
\newblock \bibinfo{journal}{\emph{arXiv preprint arXiv:2201.06850}} (\bibinfo{year}{2022}).
\newblock


\bibitem[Tufano et~al\mbox{.}(2022b)]%
        {tufano:icse22}
\bibfield{author}{\bibinfo{person}{Rosalia Tufano}, \bibinfo{person}{Simone Scalabrino}, \bibinfo{person}{Luca Pascarella}, \bibinfo{person}{Emad Aghajani}, \bibinfo{person}{Rocco Oliveto}, {and} \bibinfo{person}{Gabriele Bavota}.} \bibinfo{year}{2022}\natexlab{b}.
\newblock \showarticletitle{Using reinforcement learning for load testing of video games}. In \bibinfo{booktitle}{\emph{Proceedings of the 44th International Conference on Software Engineering}} (Pittsburgh, Pennsylvania) \emph{(\bibinfo{series}{ICSE '22})}. \bibinfo{pages}{2303–2314}.
\newblock
\showISBNx{9781450392211}
\urldef\tempurl%
\url{https://doi.org/10.1145/3510003.3510625}
\showDOI{\tempurl}


\bibitem[Vinyals et~al\mbox{.}(2017)]%
        {Vinyals2017}
\bibfield{author}{\bibinfo{person}{Oriol Vinyals}, \bibinfo{person}{Timo Ewalds}, \bibinfo{person}{Sergey Bartunov}, \bibinfo{person}{P. Georgiev}, \bibinfo{person}{A.~S. Vezhnevets}, \bibinfo{person}{Michelle Yeo}, \bibinfo{person}{Alireza Makhzani}, \bibinfo{person}{Heinrich K{\"u}ttler}, \bibinfo{person}{J. Agapiou}, \bibinfo{person}{Julian Schrittwieser}, \bibinfo{person}{John Quan}, \bibinfo{person}{Stephen Gaffney}, \bibinfo{person}{S. Petersen}, \bibinfo{person}{K. Simonyan}, \bibinfo{person}{T. Schaul}, \bibinfo{person}{H.~V. Hasselt}, \bibinfo{person}{D. Silver}, \bibinfo{person}{T. Lillicrap}, \bibinfo{person}{Kevin Calderone}, \bibinfo{person}{Paul Keet}, \bibinfo{person}{Anthony Brunasso}, \bibinfo{person}{D. Lawrence}, \bibinfo{person}{Anders Ekermo}, \bibinfo{person}{J. Repp}, {and} \bibinfo{person}{Rodney Tsing}.} \bibinfo{year}{2017}\natexlab{}.
\newblock \showarticletitle{StarCraft II: A New Challenge for Reinforcement Learning}.
\newblock \bibinfo{journal}{\emph{ArXiv}}  \bibinfo{volume}{abs/1708.04782} (\bibinfo{year}{2017}).
\newblock


\bibitem[Vuong and Takada(2018)]%
        {vuong:2018}
\bibfield{author}{\bibinfo{person}{Thi Anh~Tuyet Vuong} {and} \bibinfo{person}{Shingo Takada}.} \bibinfo{year}{2018}\natexlab{}.
\newblock \showarticletitle{A reinforcement learning based approach to automated testing of android applications}. In \bibinfo{booktitle}{\emph{Proceedings of the 9th ACM SIGSOFT International Workshop on Automating TEST Case Design, Selection, and Evaluation}}. \bibinfo{pages}{31--37}.
\newblock


\bibitem[Wan et~al\mbox{.}(2018)]%
        {wan:ase2018}
\bibfield{author}{\bibinfo{person}{Yao Wan}, \bibinfo{person}{Zhou Zhao}, \bibinfo{person}{Min Yang}, \bibinfo{person}{Guandong Xu}, \bibinfo{person}{Haochao Ying}, \bibinfo{person}{Jian Wu}, {and} \bibinfo{person}{Philip~S. Yu}.} \bibinfo{year}{2018}\natexlab{}.
\newblock \showarticletitle{Improving automatic source code summarization via deep reinforcement learning}. In \bibinfo{booktitle}{\emph{Proceedings of the 33rd ACM/IEEE International Conference on Automated Software Engineering}} \emph{(\bibinfo{series}{ASE '18})}. \bibinfo{pages}{397–407}.
\newblock
\urldef\tempurl%
\url{https://doi.org/10.1145/3238147.3238206}
\showDOI{\tempurl}


\bibitem[Wang et~al\mbox{.}(2022a)]%
        {wang:cc2022}
\bibfield{author}{\bibinfo{person}{Huanting Wang}, \bibinfo{person}{Zhanyong Tang}, \bibinfo{person}{Cheng Zhang}, \bibinfo{person}{Jiaqi Zhao}, \bibinfo{person}{Chris Cummins}, \bibinfo{person}{Hugh Leather}, {and} \bibinfo{person}{Zheng Wang}.} \bibinfo{year}{2022}\natexlab{a}.
\newblock \showarticletitle{Automating reinforcement learning architecture design for code optimization}. In \bibinfo{booktitle}{\emph{Proceedings of the 31st ACM SIGPLAN International Conference on Compiler Construction}}. \bibinfo{pages}{129–143}.
\newblock
\urldef\tempurl%
\url{https://doi.org/10.1145/3497776.3517769}
\showDOI{\tempurl}


\bibitem[Wang et~al\mbox{.}(2022b)]%
        {wang:tse22}
\bibfield{author}{\bibinfo{person}{Wenhua Wang}, \bibinfo{person}{Yuqun Zhang}, \bibinfo{person}{Yulei Sui}, \bibinfo{person}{Yao Wan}, \bibinfo{person}{Zhou Zhao}, \bibinfo{person}{Jian Wu}, \bibinfo{person}{Philip~S. Yu}, {and} \bibinfo{person}{Guandong Xu}.} \bibinfo{year}{2022}\natexlab{b}.
\newblock \showarticletitle{Reinforcement-Learning-Guided Source Code Summarization Using Hierarchical Attention}.
\newblock \bibinfo{journal}{\emph{IEEE Transactions on Software Engineering}} \bibinfo{volume}{48}, \bibinfo{number}{1} (\bibinfo{year}{2022}), \bibinfo{pages}{102--119}.
\newblock
\urldef\tempurl%
\url{https://doi.org/10.1109/TSE.2020.2979701}
\showDOI{\tempurl}


\bibitem[Weyssow et~al\mbox{.}(2024)]%
        {weyssow2024codeultrafeedback}
\bibfield{author}{\bibinfo{person}{Martin Weyssow}, \bibinfo{person}{Aton Kamanda}, {and} \bibinfo{person}{Houari Sahraoui}.} \bibinfo{year}{2024}\natexlab{}.
\newblock \showarticletitle{CodeUltraFeedback: An LLM-as-a-Judge Dataset for Aligning Large Language Models to Coding Preferences}.
\newblock \bibinfo{journal}{\emph{arXiv preprint arXiv:2403.09032}} (\bibinfo{year}{2024}).
\newblock


\bibitem[Wu et~al\mbox{.}(2020)]%
        {wu:icsme2020}
\bibfield{author}{\bibinfo{person}{Yuechen Wu}, \bibinfo{person}{Yingfeng Chen}, \bibinfo{person}{Xiaofei Xie}, \bibinfo{person}{Bing Yu}, \bibinfo{person}{Changjie Fan}, {and} \bibinfo{person}{Lei Ma}.} \bibinfo{year}{2020}\natexlab{}.
\newblock \showarticletitle{Regression Testing of Massively Multiplayer Online Role-Playing Games}. In \bibinfo{booktitle}{\emph{2020 IEEE International Conference on Software Maintenance and Evolution (ICSME)}}. IEEE, \bibinfo{pages}{692--696}.
\newblock


\bibitem[Zheng et~al\mbox{.}(2024)]%
        {zheng2024judging}
\bibfield{author}{\bibinfo{person}{Lianmin Zheng}, \bibinfo{person}{Wei-Lin Chiang}, \bibinfo{person}{Ying Sheng}, \bibinfo{person}{Siyuan Zhuang}, \bibinfo{person}{Zhanghao Wu}, \bibinfo{person}{Yonghao Zhuang}, \bibinfo{person}{Zi Lin}, \bibinfo{person}{Zhuohan Li}, \bibinfo{person}{Dacheng Li}, \bibinfo{person}{Eric Xing}, {et~al\mbox{.}}} \bibinfo{year}{2024}\natexlab{}.
\newblock \showarticletitle{Judging llm-as-a-judge with mt-bench and chatbot arena}.
\newblock \bibinfo{journal}{\emph{Advances in Neural Information Processing Systems}}  \bibinfo{volume}{36} (\bibinfo{year}{2024}).
\newblock


\bibitem[Zheng et~al\mbox{.}(2019)]%
        {zheng:ase2019}
\bibfield{author}{\bibinfo{person}{Yan Zheng}, \bibinfo{person}{Xiaofei Xie}, \bibinfo{person}{Ting Su}, \bibinfo{person}{Lei Ma}, \bibinfo{person}{Jianye Hao}, \bibinfo{person}{Zhaopeng Meng}, \bibinfo{person}{Yang Liu}, \bibinfo{person}{Ruimin Shen}, \bibinfo{person}{Yingfeng Chen}, {and} \bibinfo{person}{Changjie Fan}.} \bibinfo{year}{2019}\natexlab{}.
\newblock \showarticletitle{Wuji: Automatic online combat game testing using evolutionary deep reinforcement learning}. In \bibinfo{booktitle}{\emph{2019 34th IEEE/ACM International Conference on Automated Software Engineering (ASE)}}. IEEE, \bibinfo{pages}{772--784}.
\newblock


\bibitem[Zhu et~al\mbox{.}(2022)]%
        {zhu2022multilingual}
\bibfield{author}{\bibinfo{person}{Ming Zhu}, \bibinfo{person}{Karthik Suresh}, {and} \bibinfo{person}{Chandan~K Reddy}.} \bibinfo{year}{2022}\natexlab{}.
\newblock \showarticletitle{Multilingual code snippets training for program translation}. In \bibinfo{booktitle}{\emph{Proceedings of the AAAI Conference on Artificial Intelligence}}, Vol.~\bibinfo{volume}{36}. \bibinfo{pages}{11783--11790}.
\newblock


\bibitem[Ziegler et~al\mbox{.}(2019)]%
        {ziegler2019fine}
\bibfield{author}{\bibinfo{person}{Daniel~M Ziegler}, \bibinfo{person}{Nisan Stiennon}, \bibinfo{person}{Jeffrey Wu}, \bibinfo{person}{Tom~B Brown}, \bibinfo{person}{Alec Radford}, \bibinfo{person}{Dario Amodei}, \bibinfo{person}{Paul Christiano}, {and} \bibinfo{person}{Geoffrey Irving}.} \bibinfo{year}{2019}\natexlab{}.
\newblock \showarticletitle{Fine-tuning language models from human preferences}.
\newblock \bibinfo{journal}{\emph{arXiv preprint arXiv:1909.08593}} (\bibinfo{year}{2019}).
\newblock


\end{thebibliography}

\end{document}